# Open model-based analysis of a 100% renewable and sector-coupled energy system–The case of Germany in 2050

Md. Nasimul Islam Maruf[*]

*Department of Energy and Environmental Management, Europa-Universität Flensburg, Flensburg, Germany*

HIGHLIGHTS

- Open model-based analysis for 100% renewable and sector-coupled energy systems.
- Cost-optimization tool based on Oemof to investigate Germany's energy system in 2050.
- Renewable energy is sufficient for electricity and building heat in Germany in 2050.
- Investment cost (bn €/yr): 17.6–26.6 (Volatile), 23.7–28.8 (Heat), 2.7–3.9 (Storage).
- Energy mix and LCOE comparison validate the developed tool's effectiveness.

GRAPHICAL ABSTRACT

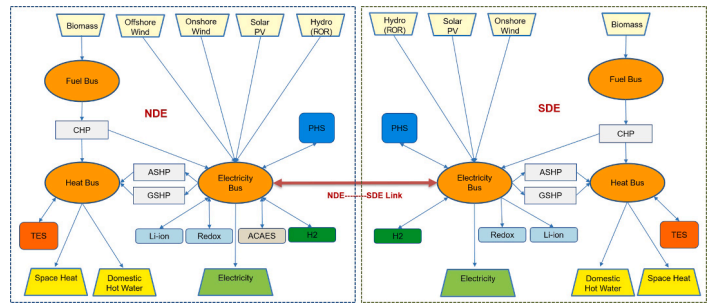

ABSTRACT

The ambitious energy target to achieve climate-neutrality in the European Union (EU) energy system raises the feasibility question of using only renewables across all energy sectors. As one of the EU's leading industrialized countries, Germany has adopted several climate-action plans for the realistic implementation and maximum utilization of renewable energies in its energy system. The literature review shows a clear gap in comprehensive techniques describing an open modeling approach for analyzing fully renewable and sector-coupled energy systems. This paper outlines a method for analyzing the 100% renewable-based and sector-coupled energy system's feasibility in Germany. Based on the open energy modeling framework, an hourly optimization tool 'OSeEM-DE' is developed to investigate the German energy system. The model results show that a 100% renewable-based and sector-coupled system for electricity and building heat is feasible in Germany. The investment capacities and component costs depend on the parametric variations of the developed scenarios. The annual investment costs vary between 17.6 and 26.6 bn €/yr for volatile generators and between 23.7 and 28.8 bn €/yr for heat generators. The model suggests an investment of a minimum of 2.7–3.9 bn €/yr for electricity and heat storage. Comparison of OSeEM-DE results with recent studies validates the percentage-wise energy mix composition and the calculated Levelized Cost of Electricity (LCOE) values from the model. Sensitivity analyses indicate that storage and grid expansion maximize the system's flexibility and decrease the investment cost. The study concludes by showing how the tool can analyze different energy systems in the EU context.

* Address: Auf dem Campus 1b (Room: VIL 108), 24943 Flensburg, Germany.
  *E-mail address:* ni.maruf@uni-flensburg.de.

**Nomenclature**

*Abbreviations*

| | |
|---|---|
| ACAES | Adiabatic Compressed Air Energy Storage |
| ASHP | Air Source Heat Pump |
| Capex | Capital Expenditure |
| CHP | Combined Heat and Power |
| CSP | Concentrated Solar Power |
| DE | Germany |
| Dispa-SET | Unit commitment and optimal dispatch model |
| DSM | Demand-side Management |
| ENTSO-e | European Network of Transmission System Operators for Electricity |
| EU | European Union |
| FOM | Fixed Operation and Maintenance |
| GHG | Greenhouse Gas |
| GSHP | Ground Source Heat Pump |
| $H_2$ | Hydrogen |
| HVDC | High Voltage Direct Current |
| ICT | Information and Communications Technology |
| LCOE | Levelized Cost of Electricity |
| Li-ion | Lithium-ion |
| LP | Linear Programming |
| MERRA | Modern-Era Retrospective analysis for Research and Application |
| MILP | Mixed-Integer Linear Programming |
| NDE | Northern Germany |
| NPV | Net Present Value |
| NS | North Sea |
| O&M | Operation and Maintenance |
| Oemof | Open Energy Modelling Framework |
| OPSD | Open Power System Data |
| OSeEM-DE | Open Sector-coupled Energy Model for Germany |
| P2H | Power-to-Heat |
| P2G | Power-to-Gas |
| PHS | Pumped Hydro Storage |
| PV | Photovoltaic |
| PyPSA | Python for Power System Analysis |
| Redox | Vanadium Redox |
| ROR | Run-of-the-River |
| SDE | Southern Germany |
| TES | Thermal Energy Storage |
| V2G | Vehicle-to-Grid |
| VOM | Variable Operation and Maintenance |
| WACC | Weighted Average Cost of Capital |
| WWS | Wind-Water-Solar |
| ZNES | Center for Sustainable Energy Systems |

*Symbols*

| | |
|---|---|
| $t$ | Timestep |
| $b$ | Bus |
| $l$ | Load |
| $v$ | Volatile generator |
| $h$ | Heat pump |
| $s$ | Storage |
| $m$ | Transmission line |
| $T$ | Set of all timesteps |
| $B$ | Set of all buses |
| $L$ | Set of all loads |
| $V$ | Set of all volatile generators |
| $H$ | Set of all heat pumps |
| $S$ | Set of all storages |
| $M$ | Set of all transmission lines |

*Variables*

| | |
|---|---|
| $x_{b,in}^{flow}$ | Input flow to bus $b$ |
| $x_{b,out}^{flow}$ | Output flow from bus $b$ |
| $x_l^{flow}$ | Load flow of load $l$ |
| $x_v^{flow}$ | Flow of volatile generator unit $v$ |
| $x_v^{capacity}$ | Endogenous capacity of volatile generator unit $v$ |
| $x_{chp}^{flow,carrier}$ | Carrier flow of CHP unit $chp$ |
| $x_{chp}^{flow,electricity}$ | Electricity flow of CHP unit $chp$ |
| $x_{chp}^{flow,heat}$ | Heat flow of CHP unit $chp$ |
| $x_{h,from}^{flow}$ | Electricity flow from the electricity bus for heat pump unit $h$ |
| $x_{h,to}^{flow}$ | Heat flow to the heat bus for heat pump unit $h$ |
| $x_s^{level}$ | Energy level of storage unit $s$ |
| $x_{s,in}^{flow}$ | Input flow to storage unit $s$ |
| $x_{s,out}^{flow}$ | Output flow from storage unit $s$ |
| $x_{phs}^{level}$ | Energy level of pumped hydro storage unit $phs$ |
| $x_{phs}^{flow,out}$ | Output flow of pumped hydro storage unit $phs$ |
| $x_{phs}^{profile}$ | Endogenous inflow profile of pumped hydro storage unit $phs$ |
| $x_{m,from}^{flow}$ | Flow from a bus through the transmission line $m$ |
| $x_{m,to}^{flow}$ | Flow to a bus through the transmission line $m$ |

*Parameters*

| | |
|---|---|
| $c_l^{profile}$ | Exogenous profile of load $l$ |
| $c_l^{amount}$ | Total amount of load $l$ |
| $c_v^{profile}$ | Exogenous profile of volatile generator $v$ |
| $c_v^{capacity\_potential}$ | Maximum capacity potential of volatile generator $v$ |
| $c_{chp}^{beta}$ | Power loss index of CHP |
| $c_{chp}^{electrical\_efficiency}$ | Electrical efficiency of CHP |
| $c_{chp}^{thermal\_efficiency}$ | Thermal efficiency of CHP |
| $c_{chp}^{condensing\_efficiency}$ | Condensing efficiency of CHP |
| $c_{biomass}^{amount}$ | Absolute amount of biomass commodity |
| $c_h^{efficiency}$ | Conversion efficiency (Coefficient of Performance, i.e., COP) of heat pump unit $h$ |
| $c_s^{loss\_rate}$ | Loss rate of storage unit $s$ |
| $c_s^{eta\_in}$ | Charging efficiency of storage unit $s$ |
| $c_s^{eta\_out}$ | Discharging efficiency of storage unit $s$ |
| $c_s^{roundtrip\_efficiency}$ | Round Trip efficiency of storage unit $s$ |
| $c_s^{capacity}$ | Maximum power capacity of storage unit $s$ |
| $c_s^{storage\_capacity}$ | Maximum energy capacity of storage unit $s$ |
| $c_{phs}^{loss\_rate}$ | Loss rate of pumped hydro storage unit $phs$ |
| $c_{phs}^{efficiency}$ | Efficiency of pumped hydro storage unit $phs$ |
| $c_{phs}^{profile}$ | Exogenous inflow profile of pumped hydro storage unit $phs$ |
| $c_m^{loss}$ | Loss on transmission line $m$ |
| $c^{marginal\_cost}$ | Marginal cost |
| $c^{VOM}$ | Variable operation and maintenance cost |
| $c^{carrier\_cost}$ | Carrier cost |
| $\eta$ | Efficiency |
| $c^{capacity\_cost}$ | Capacity cost |
| $c_s^{storage\_capacity\_cost}$ | Energy capacity cost of storage |
| $c^{annuity}$ | Annuity cost |
| $c^{capex}$ | Capital expenditure cost |
| $c^{WACC}$ | Weighted Average Cost of Capital |
| $c^{FOM}$ | Fixed operation and maintenance Cost |

| | | | |
|---|---|---|---|
| $n$ | Lifetime | $I_n$ | Initial cost of investment expenditure |
| $c^{AIV}$ | Annual investment cost | $M_n$ | Sum of all O&M costs |
| $c^{optimized\_capacity}$ | Optimized capacity cost | $F_n$ | Sum of all fuel Costs |
| $c^{TIV}$ | Total Investment cost | $E_n$ | Sum of all electrical energy produced |
| $c^{LCOE}$ | Levelized cost of electricity | $r$ | Discount rate |

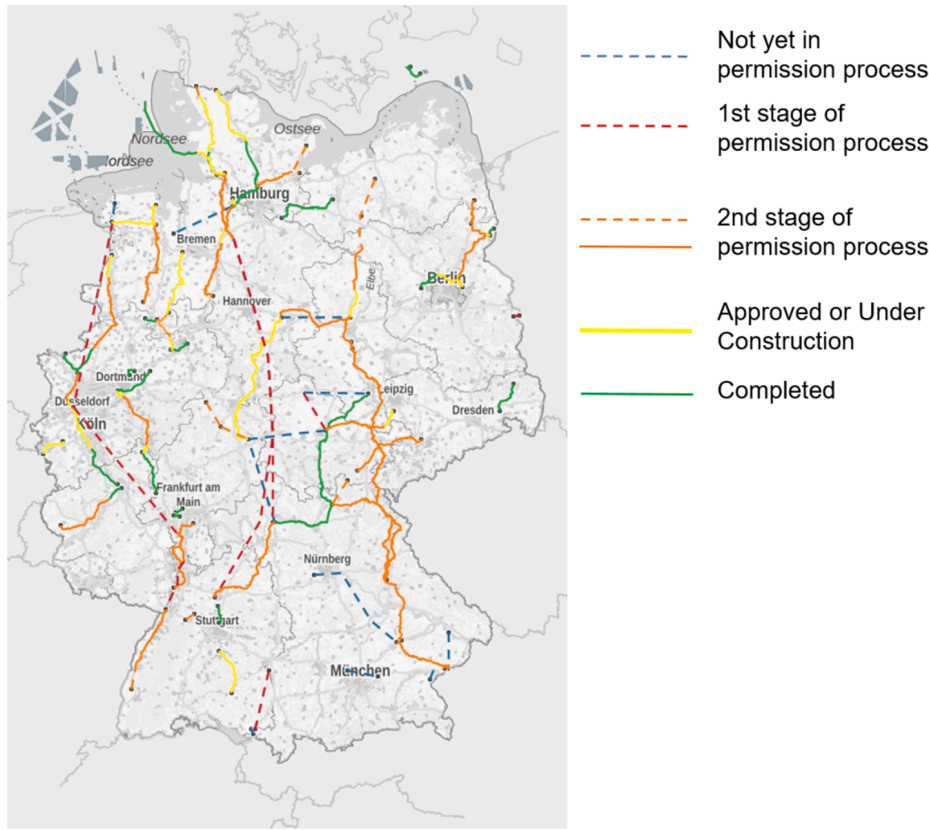

**Fig. 1.** Current status of electricity transmission network planning and development in Germany. The orange, red, and yellow lines show that many North-South grid extension projects are in the planning and development phase. Source: Federal Network Agency of Germany (Bundesnetzagentur) 2020 [9].

## 1. Introduction

A clear rulebook was adopted for the practical implementation of the Paris Agreement in December 2018 at the United Nations Climate Conference (COP24) in Katowice, which delineated a world-wide climate action plan to mitigate climate change by reducing global warming to 1.5 °C if possible [1]. To achieve a climate-neutral society with net-zero greenhouse gas (GHG) emissions across all sectors by 2050, the European Commission endorsed a strategic long-term vision in line with the Paris Agreement for the European Union (EU) [2]. The North Sea (NS) region is a representative area to demonstrate the feasibility and plausible pathways for achieving the ambitious target of net-zero emission across all the sectors in an energy system [3]. As a leading industrialized nation of the NS region and the EU, Germany already adopted a climate action plan for the realistic implementation of extensive GHG neutrality across all the energy sectors by 2050 [4]. This ambitious energy target requires the 'reinvention' of energy systems to be technically feasible and economically viable, guaranteeing realistic solutions to ensure affordability, reliability, and sustainability [5].

The overcapacity of volatile generators in a 100% renewable-based energy system leads to excess energy in the electricity grid. A relatively new approach known as 'sector coupling' can efficiently use this surplus energy [6]. Sector coupling has emerged as a new concept in energy and climate policy discussions in recent years. The 'Climate Action Plan 2050′ and the 'Green Paper on Energy Efficiency' reflect the high political significance of sector coupling, among other things in Germany [4,7]. Sector coupling can make a decisive contribution to the achievement of ambitious climate protection goals through increased use of renewable electricity in the heating and transport sectors and industry to substitute fossil fuels.

The expansion of renewable energies in Germany will increase the need for grid expansion, particularly in local distribution grids for solar photovoltaic (PV) and onshore wind plants, and in transmission grids for offshore wind power. The current power system of Germany has many fossil fuel-based and nuclear-based power plants in Southern Germany. A gradual exit from fossil fuel and nuclear power will require shifting to renewables from offshore wind plants in Northern Germany. Therefore, there will be a need for transmission grid expansion from North to South at the same time. The transmission grid operators and the federal government of Germany have accelerated the development of Grids from Northern to Southern Germany, with 4,650 km to be constructed by 2025 [8]. Fig. 1 shows the status of grid expansion in Germany in which the expansion plan from north to south and other grids in the planning and development stages is illustrated [9].

Flexibility and sector coupling can bring renewable energies into the various energy sectors in high proportions and effectively increase energy efficiency potentials. By adopting solutions like power-to-heat (P2H) technologies (such as electric heat pumps), and power-to-gas (P2G) technologies (such as Hydrogen storage via electrolysis), the transformation of energy between different energy sectors can be made even more independent of the supply and demand. Flexibility can be offered in various ways, for example, shifting heat pumps or electric vehicle loads in private households and even shifting the energy consumption in industries as a part of a sector-coupled network. However, the energy flow should be smartly and effectively managed between the sectors to stabilize the network operation while keeping the system flexible and efficient [10].

## 2. State of the art and research question

In the last two decades, both 'sector coupling' and '100% renewable energy systems' have been subjects of interest for energy researchers globally, as reviewed by the author in an earlier article [3]. Several pieces of research investigated 100% or near-100% renewable energy systems from national perspectives. Such investigation includes energy system analysis of Australia [11,12], Barbados [13], Belgium [14], Brazil [15–18], Canada [17], China [19], Colombia [20], Costa Rica [17], Croatia [21], Denmark [22–26], Finland [26,27], France [28], Germany [29–31], Great Britain [32], Iceland [26], India and the SAARC region [33,34], Iran [35], Ireland [36,37], Italy [38], Japan [39], Macedonia [40], New Zealand [41], Nicaragua [42], Nigeria [43], Norway [17,26], Pakistan [44], Paraguay [5,18], Portugal [45], Saudi Arabia [46], Seychelles [47], Tokelau [48], Turkey [49], Ukraine [50], the United Kingdom [51–53], the United States [54–56], and Uruguay [18]. Other than these national studies, there are many other 100% renewable system studies larger than national energy systems covering the World [55,57–64], North-East Asia [65], the ASEAN region [66], Europe and its neighbors [67], Europe [68–72], South-East Europe [73], and the Americas [74]. Similarly, there are also a number of regional-level studies on 100% system including Hvar (Croatia) [75], Samsø (Denmark) [76], the Åland Islands (Finland) [77], Mecklenburg-Vorpommern (Germany) [78], Schleswig-Holstein (Germany) [78], Orkney (Scotland) [76], the Canary Islands (Spain) [79], California (USA) [80], and New York State (USA) [81]. Most of these studies focus on 100% renewable-based electricity systems, and only a few consider energy transition pathways for reaching the target, including all relevant energy sectors [82]. Among the 100% renewable energy system studies that include all energy sectors, there are two main technologically and economically feasible concepts. The first concept is 'Smart Energy Systems,' which has been studied and analyzed in many research articles, including [21–25,31,36,37,71]. The concept takes an integrated and holistic focus to include electricity, heating, cooling, industry, building, and transportation sectors and discusses the potential benefits of sectoral and infrastructure integration in the energy system. The second concept is 'wind-water-solar (WWS) for all purposes', which has been discussed in many research articles, including [55,56,59–64,80,81]. The WWS concept presents roadmaps from local to global levels to electrify all energy sectors (i.e., transportation, heating, cooling, industry, agriculture, forestry, and fishing) using only wind, water, and solar power. The WWS roadmaps present the benefits of the energy transition in terms of energy access, mitigation of global warming and avoidance of air pollution deaths, creation of jobs, and costs such as energy, health, climate, and social costs.

In German energy system analyses, several studies focus on individual energy sectors such as the heating market, transportation, future electricity market. For achieving the energy transition towards climate-neutrality, Schmid et al. identified and characterized actors who can put the energy transition into practice, but they focused on the power sector only [83]. In a similar publication, Lehmann and Nowakowski analyzed three scenarios for Germany's future electricity system [84]. Robinius et al. reviewed sector coupling scenarios for Germany linking electricity and transport sectors [6,85]. Gullberg et al. described how Norway and Germany's interconnection could ensure a low-carbon energy future [86]. Scholz et al. identified the bottlenecks of using only renewables for the energy transition [87]. Schroeder et al. compared different scenarios to investigate the need for grid expansions in Germany [88].

Although there are various models and analysis on the German energy transition towards climate neutrality, most of them focus either on specific aspects of an individual energy sector or the markets and actors linked with the energy system. A few studies outline a detailed techno-economic analysis using 100% renewables, including multiple energy sectors. Henning and Palzer showed that a 100% renewable energy system for power and heat is technically possible, and the overall annual cost for the system will be comparable to today's price [29,30]. However, their analysis assumes that heat requirement is reduced by 60% in the building sector in the future compared to today's demand, and their analyzing tool REMod-D is not an open-source model. In another recent publication, Hansen et al. concluded that the full energy system transition towards 100% renewables by 2050 in Germany is technically and economically possible, but resource potentials such as biomass is a big challenge for this transition [31]. Hansen et al. used EnergyPLAN for their analysis, a popular free-to-use simulation tool for modeling 100% systems for all energy sectors. However, it is not possible to analyze the capacity expansion mechanism and reach an optimum solution using EnergyPLAN [89]. A literature review by the author shows a clear research gap in applying cross-sectoral holistic approaches using comprehensive open modeling techniques for analyzing 100% renewable and sector-coupled energy systems in Germany and the other countries in the NS region [3]. Therefore, based on the scope, the following research question is formulated–

*How feasible is it to transform towards a 100% renewable energy system for both power and building heat in Germany by 2050?*

Answering the main research question raises additional questions on the overall energy mix, investment cost and capacities, the issue of grid expansion within Germany, and different flexibility aspects of the system; which formulates the following sub-research questions–

1. *In which capacities could such a system be implemented?*
2. *What are the estimated investment costs for the different system components?*
3. *What are the flexibility aspects of grid expansion, storage, and dispatchable loads in such a system?*

The paper's main objective is to design and develop an open-source energy model and investigate the German power and heat sectors. The model analyzes different scenarios to investigate moderate to extreme conditions to determine the plausible energy mix and required investment in component-wise capacities and costs. The model considers dividing the German energy system into two sub-national regions to analyze the grid expansion from Northern to Southern Germany. It also helps to understand the underlying energy flow mechanism and different energy components' roles from national and sub-national perspectives. The heating sector analysis in this model is limited to the use of building heat, including space heating and domestic hot water demands. The present study does not consider the industrial process heating demands. The industrial process heating is planned to be included in the upgraded version of the model.

Section 3 describes the model's methodology, which consists of the model's architecture and its component-wise mathematical description. Section 4 describes the details of the developed model's working mechanism and the input data used for the analysis. Section 5 describes the development of different scenarios. Section 6 presents and compares the scenario results and examines the system's details based on the research questions. Section 7 summarizes the results, discusses the limitations, model development plans, and future steps of the study, and concludes with the model's usability in a broader context.

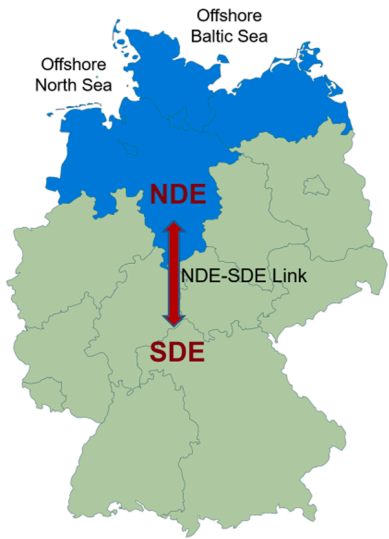

**Fig. 2.** The German national energy system with two sub-national nodes, NDE (shown using blue color) and SDE (shown using green color). Map source: Wikimedia Commons [91].

## 3. Methodology

The optimization model follows a hybrid approach where the current technology capacities are exogenously defined, and the future investment capacities are endogenously determined. System boundaries, according to technology potentials, are set to avoid any unrealistic solution based on overestimation. The 'Open Sector-coupled Energy Model for Germany (OSeEM-DE)' is an application constructed using 'Oemof Tabular,' which is a python-based linear programming (LP) optimization tool under the open energy modeling framework (Oemof) [90].

As shown in Fig. 2, the two German sub-national energy system nodes are Northern Germany (NDE) and Southern Germany (SDE). NDE consists of Schleswig-Holstein, Lower Saxony, Bremen, Hamburg, and Mecklenburg-Vorpommern, and SDE consists of the remaining eleven states of Germany. The power exchange between NDE and SDE is possible using the transshipment approach via NDE-SDE Link.

Fig. 3 illustrates the Oemof-based OSeEM-DE energy system model. The model's volatile components are Onshore wind, Offshore wind, Solar PV, and Hydro Run-of-the-River (ROR) plants. The fuel input of the Combined Heat and Power (CHP) plants comes from biomass resources. In addition to CHPs, two types of heat pumps are used to supply heat loads in the model: Air Source Heat Pump (ASHP) and Ground Source Heat Pump (GSHP). The state-of-the-art GSHPs are becoming cost-competitive compared to ASHPs and offer a more energy-efficient solution because of their use of consistent ground temperatures. The storage options are Pumped Hydro Storage (PHS), Lithium-ion (Li-ion) battery, Vanadium Redox (Redox) flow battery, Hydrogen ($H_2$) storage, and Adiabatic Compressed Air Energy Storage (ACAES). The heat storage option is Thermal Energy Storage (TES) using hot water tanks. NDE and SDE energy systems are almost identical, except that Offshore wind and ACAES components are only available in NDE.

The OSeEM-DE model uses 'Oemof Solph' [92], following the formulation described by Hilpert in [93], where the maximum potential constrains the volatile renewable sources, total biomass amount, and storage capacities. The transmission capacity between NDE and SDE and the heat storage capacities are also limited exogenously. The upper limit for the P2H technologies are subject to optimization but depends on the availability of electricity. The model uses a perfect foresight approach; hence all timesteps of the model time horizon are read by the solver at once. Therefore, the weather and the renewable supply data for the full year is known in advance. The different components of the energy system and their mathematical formulations are as follows, where $x$ denotes the endogenous variables, and $c$ denotes the exogenous parameters used in the model.

### 3.1. Bus

There are three types of buses in the energy model: electrical, heat, and fuel bus. Since all flows into and out of a bus are balanced, for the set of all buses $b \in B$, sum of all input flows $x^{flow}_{b,in}$ to a bus $b$ must be equal to the sum of all output flows $x^{flow}_{b,out}$:

$$\sum x^{flow}_{b,in}(t) = \sum x^{flow}_{b,out}(t) \quad \forall t \in T, \forall b \in B \qquad (1)$$

### 3.2. Load

Two types of loads are considered in the energy model: electricity and heat. The heat load consists of space heat and domestic hot water. For the set of all loads $l \in L$, the load flow $x^{flow}_l$ at every time step $t$ must be equal to the product of two exogenously given inputs: profile value of the load $c^{profile}_l$ at time step $t$, and the total amount of the load $c^{amount}_l$:

$$x^{flow}_l(t) = c^{profile}_l(t) \cdot c^{amount}_l \quad \forall t \in T, \forall l \in L \qquad (2)$$

### 3.3. Volatile generator

The volatile generators in the model are onshore and offshore wind, solar PV, and hydro ROR plants. These components are connected to electricity buses. The offshore wind component is available only in NDE. For all the volatile generators $v \in V$, the flow $x_v^{flow}$ at every time step $t$ must be equal to the product of the capacity $x_v^{capacity}$ and the profile value of the volatile generator $c_v^{profile}$:

$$x_v^{flow}(t) = x_v^{capacity}(t) \cdot c_v^{profile}(t) \quad \forall t \in T, \forall v \in V \tag{3}$$

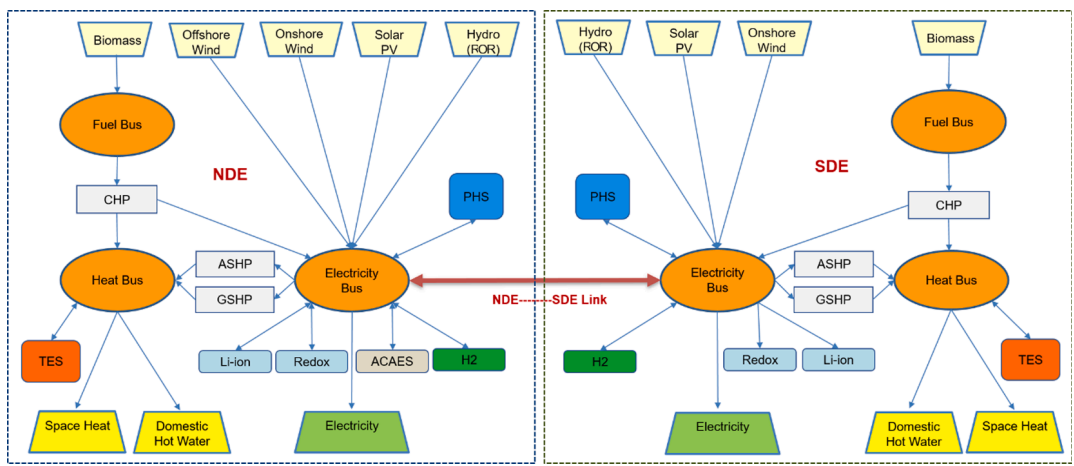

**Fig. 3.** Simplified block diagram of the OSeEM-DE model for Germany showing NDE and SDE nodes.

The endogenously obtained capacity of the generator $x_v^{capacity}$ at every time step $t$ must be less than or equal to the maximum capacity potential of the volatile component $c_v^{capacity\_potential}$:

$$x_v^{capacity}(t) \leq c_v^{capacity\_potential} \quad \forall t \in T, \forall v \in V \tag{4}$$

### 3.4. Combined heat and power (CHP)

Since the model is 100% RES-based, the CHP plants use only biomass resources as fuel. Thus, biomass goes into the fuel bus, which later goes as the input of CHP, as shown in Fig. 3. The CHP runs in extraction turbine mode, as shown by (5–7), according to Mollenhauer et al. [94]. (5) shows the relation between the input carrier flow $x_{chp}^{flow,carrier}$ and the two output flows: electrical output flow $x_{chp}^{flow,electricity}$ and heat output flow $x_{chp}^{flow,heat}$, at every time step $t$. (6) shows the relationship between the two output flows. $c_{chp}^{beta}$ is the power loss index, which is derived using (7). $c_{chp}^{electrical\_efficiency}$, $c_{chp}^{thermal\_efficiency}$ and $c_{chp}^{condensing\_efficiency}$ denote the electrical, thermal and condensing efficiencies of the CHP unit, respectively.

$$x_{chp}^{flow,carrier}(t) = \frac{x_{chp}^{flow,electricity}(t) + x_{chp}^{flow,heat}(t) \cdot c_{chp}^{beta}}{c_{chp}^{condensing\_efficiency}} \quad \forall t \in T \tag{5}$$

$$x_{chp}^{flow,electricity}(t) \geq x_{chp}^{flow,heat}(t) \cdot \frac{c_{chp}^{electrical\_efficiency}}{c_{chp}^{thermal\_efficiency}} \quad \forall t \in T \tag{6}$$

$$c_{chp}^{beta} = \frac{c_{chp}^{condensing\_efficiency} - c_{chp}^{electrical\_efficiency}}{c_{chp}^{thermal\_efficiency}} \tag{7}$$

Equation (8) models the limited availability of biomass commodities where the aggregated inflows are constrained by the absolute amount of the biomass commodity $c_{biomass}^{amount}$:

$$\sum x_{chp}^{flow,carrier}(t) \leq c_{biomass}^{amount} \quad \forall t \in T \tag{8}$$

### 3.5. Heat pump

Oemof Solph's conversion component, which converts power to heat, is used for modeling both GSHP and ASHP. Therefore, a conversion process of one input flow $x_{h,from}^{flow}$ (electricity flow from the electricity bus) and one output $x_{h,to}^{flow}$ (heat flow to the heat bus) and a conversion factor $c_{h,to}^{efficiency}$ (conversion efficiency i.e., coefficient of performance of the heat pump) at every time step $t$ models all the heat pumps $h \in H$:

$$x_{h,to}^{flow}(t) = x_{h,from}^{flow}(t) \cdot c_{h,to}^{efficiency} \quad \forall t \in T, \forall h \in H \tag{9}$$

### 3.6. Storage

In addition to PHS, the other electricity storage technologies used in the model are Li-ion and Redox batteries, ACAES, and $H_2$ storage. Hot water-based TES is the only heat storage component in both NDE and SDE. Oemof Solph's generic storage formulations are used to model all the storage components except PHS. For all these storages $s \in S$, the mathematical model includes the input and output flows and the storage level:

$$x_s^{level}(t) = x_s^{level}(t-1) \cdot \left(1 - c_s^{loss\_rate}\right) + c_s^{eta\_in} \cdot x_s^{flow,in}(t) - \frac{x_s^{flow,out}(t)}{c_s^{eta\_out}} \quad \forall t \in T, \forall s \in S \tag{10}$$

where $x_s^{level}$ indicates the storage energy level, $c_s^{loss\_rate}$ marks the loss rate for the storage, $x_s^{flow,in}$ and $x_s^{flow,out}$ indicates the input and output flows, and $c_s^{eta\_in}$ and $c_s^{eta\_out}$ denotes the charging and discharging efficiencies of the storage.

The charging and discharging efficiencies are formulated from the roundtrip efficiency using $c_s^{eta} = \sqrt{c_s^{roundtrip\_efficiency}}$. The input and output flows $x_s^{flow}$ are constrained by the maximum power capacity $c_s^{capacity}$ as shown in (11). The energy level $x_s^{level}$ is constrained by the maximum energy capacity $c_s^{storage\_capacity}$ as shown in (12).

$$x_s^{flow}(t) \leq c_s^{capacity} \quad \forall t \in T, \forall s \in S \tag{11}$$

$$x_s^{flow}(t) \leq c_s^{storage\_capacity} \quad \forall t \in T, \forall s \in S \tag{12}$$

### 3.7. Pumped hydro storage

The PHS are modeled as storage units with a constant inflow and possible spillage:

$$x_{phs}^{level}(t) = x_{phs}^{level}(t-1) \cdot \left(1 - c_{phs}^{loss\_rate}\right) + x_{phs}^{profile}(t) - \frac{x_{phs}^{flow,out}(t)}{c_{phs}^{efficiency}} \quad \forall t \in T \tag{13}$$

where $x_{phs}^{level}$ indicates the pumped hydro storage energy level, $c_{phs}^{loss\_rate}$ marks the loss rate for the PHS, $x_{phs}^{profile}$ indicates the endogenous inflow profile, $x_{phs}^{flow,out}$ denotes the output flow from the PHS, and $c_{phs}^{efficiency}$ marks the efficiency of the PHS. The hydro inflow is constrained by an exogenous inflow profile:

$$0 \leq x_{phs}^{profile}(t) \leq c_{phs}^{profile}(t) \quad \forall t \in T \tag{14}$$

Therefore, if the inflow exceeds the maximum storage capacity, spillage is possible by setting $x_{phs}^{profile}$ to lower values. The spillage at time step $t$ can therefore be defined by $c_{phs}^{profile}(t) - x_{phs}^{profile}(t)$.

### 3.8. Transmission line

The electricity transmission between two energy systems via NDE-SDE Link is modeled using the transshipment approach, which facilitates simple electricity exchange (import-export) after considering a transmission loss factor. Therefore, transmission lines $m \in M$ are modeled according to the following mathematical formulation:

$$x_{m,from}^{flow}(t) = \left(1 - c_m^{loss}\right) \cdot x_{m,to}^{flow}(t) \quad \forall t \in T, \forall m \in M \tag{15}$$

where $x_{from,m}^{flow}$ indicates the flow from the supplying energy system, $x_{m,to}^{flow}$ indicates the flow to the receiving energy system, and $c_m^{loss}$ indicates the loss on transmission line $m$.

### 3.9. Costs

The marginal costs $c_{v,chp,h,s}^{marginal\_cost}$ are calculated based on the variable operation and maintenance (VOM) cost $c_{v,chp,h,s}^{VOM}$, carrier cost $c_{v,chp,h,s}^{carrier\_cost}$, and the efficiency of the corresponding technology $\eta_{v,chp,h,s}$:

$$c_{v,chp,h,s}^{marginal\_cost} = c_{v,chp,h,s}^{VOM} + \frac{c_{v,chp,h,s}^{carrier\_cost}}{\eta_{v,chp,h,s}} \tag{16}$$

The capacity costs for volatile generators, CHP and heat pumps

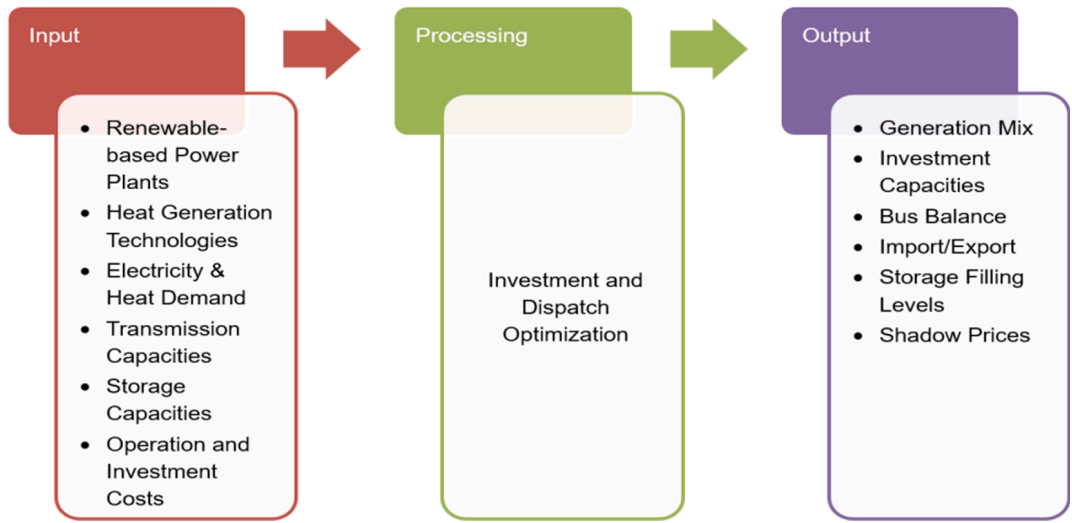

**Fig. 4.** Input and output of the OSeEM-DE model.

$c_{v,chp,h}^{capacity\_cost}$ are calculated based on the fixed operation and maintenance (FOM) cost $c_{v,chp,h}^{FOM}$, and the annuity of the corresponding technology $c_{v,chp,h}^{annuity}$, as shown in (17). The annuity $c^{annuity}$ is calculated using (18), which consist of the initial capital expenditure $c^{capex}$, the weighted average cost of capital $c^{WACC}$, and the lifetime of the investment $n$ for the respective technology.

$$c_{v,chp,h}^{capacity\_cost} = c_{v,chp,h}^{FOM} + c_{v,chp,h}^{annuity} \tag{17}$$

$$c^{annuity} = c^{capex} \cdot \frac{(c^{WACC} \cdot (1 + c^{WACC})^n}{((1 + c^{WACC})^n - 1)} \tag{18}$$

In the case of storages, the capacity costs $c_s^{capacity\_cost}$ and the storage capacity costs $c_s^{storage\_capacity\_cost}$ are calculated using (19–20):

$$c_s^{capacity\_cost} = c_{s,power}^{annuity} \tag{19}$$

$$c_s^{storage\_capacity\_cost} = c_s^{FOM} + c_{s,energy}^{annuity} \tag{20}$$

where, $c_{s,power}^{annuity}$ is the annuity cost of storage power, and $c_{s,energy}^{annuity}$ is the annuity cost of storage energy. These two annuities are calculated using the formulation of (18). $c_s^{FOM}$ is the fixed operation and maintenance cost (FOM) of energy storage.

In addition to these modeling equations, some costs are calculated and analyzed after a successful model run, based on the optimization results. The annual investment cost $c^{AIV}$ is calculated multiplying the annuity $c^{annuity}$ and the optimized capacity $c^{optimized\_capacity}$, as shown in (21). The total investment (overnight) investment cost $c^{TIV}$ is calculated multiplying the capital expenditure $c^{capex}$, and the optimized capacity $c^{optimized\_capacity}$, as shown in (22). Furthermore, the Levelized Costs of Electricity $c^{LCOE}$ is calculated from the ratio of the Net Present Value (NPV) of total costs of over lifetime (including fixed and variable operation and maintenance costs, fuel costs, and the discount rate), and the NPV of the electrical energy produced over the lifetime, as shown in (23).

$$c^{AIV} = c^{annuity} \cdot c^{optimized\_capacity} \tag{21}$$

$$c^{TIV} = c^{capex} \cdot c^{optimized\_capacity} \tag{22}$$

$$c^{LCOE} = \frac{\sum \frac{(I_n + M_n + F_n)}{(1+r)^n}}{\sum \frac{E_n}{(1+r)^n}} \tag{23}$$

where $I_n$ indicates the initial cost of investment expenditures (same as $c^{capex}$), $M_n$ indicates the sum of all operation and maintenance expenditures (sum of $c^{VOM}$ and $c^{FOM}$), and $F_n$ indicates the fuel expenditure (such as the carrier cost of biomass), $E_n$ indicates the sum of all electrical energy generation, $r$ indicates the discount rate (same as $c^{WACC}$), and $n$ indicates the lifetime.

### 3.10. Objective function

The objective function is created from all instantiated objects which use all operating costs and investment costs arguments:

$$\min : \sum_{v,t}^{operating\_cost\ Volatile\ Generator} x_v^{flow}(t) \cdot c_v^{marginal\_cost} + \sum_{v}^{investment\_cost\ Volatile\ Generator} x_v^{capacity} \cdot c_v^{capacity\_cost} +$$

$$\sum_{chp,t}^{operating\_cost\ CHP} x_{chp}^{flow}(t) \cdot c_{chp}^{marginal\_cost} + \sum_{chp}^{investment\_cost\ CHP} x_{chp}^{capacity} \cdot c_{chp}^{capacity\_cost} +$$

$$\sum_{h,t}^{operating\_cost\ Heat\ Pump} x_h^{flow}(t) \cdot c_h^{marginal\_cost} + \sum_{h}^{investment\_cost\ Heat\ Pump} x_h^{capacity} \cdot c_h^{capacity\_cost} +$$

$$\sum_{s,t}^{operating\_cost\ Storage} x_s^{flow}(t) \cdot c_s^{marginal\_cost} +$$

$$\sum_{s}^{investment\_cost\ Storage} x_s^{capacity} \cdot c_s^{capacity\_cost} + x_s^{storage\_capacity} \cdot c_s^{storage\_capacity\_cost} \tag{24}$$

## 4. Model

The underlying concept of the OSeEM-DE model uses the formulation of LP and mixed-integer linear programming (MILP) from a generic object-oriented structure of 'Oemof Solph' [92]. Fig. 4 shows a summary of the OSeEM-DE model input and output.

The input data used in the model are the capacities and potentials of renewable energy-based power plants (such as wind, PV), heat generation technologies (such as CHP, heat pump), electricity and heat demands, transshipment capacity, storage capacities, and their potential. Also, operating costs (i.e., marginal cost) and the investment costs (i.e., capacity cost) for all the considered technologies are provided as input. After solving the model using open model solvers, the results are post-processed using Oemof Tabular's post-processing scripts. The output of the model provides information on the generation mix, investment capacities according to the provided capacities and available potential, balance in each of the buses, import/export between energy systems (in this case between NDE and SDE), filling level of the storages and the shadow prices.

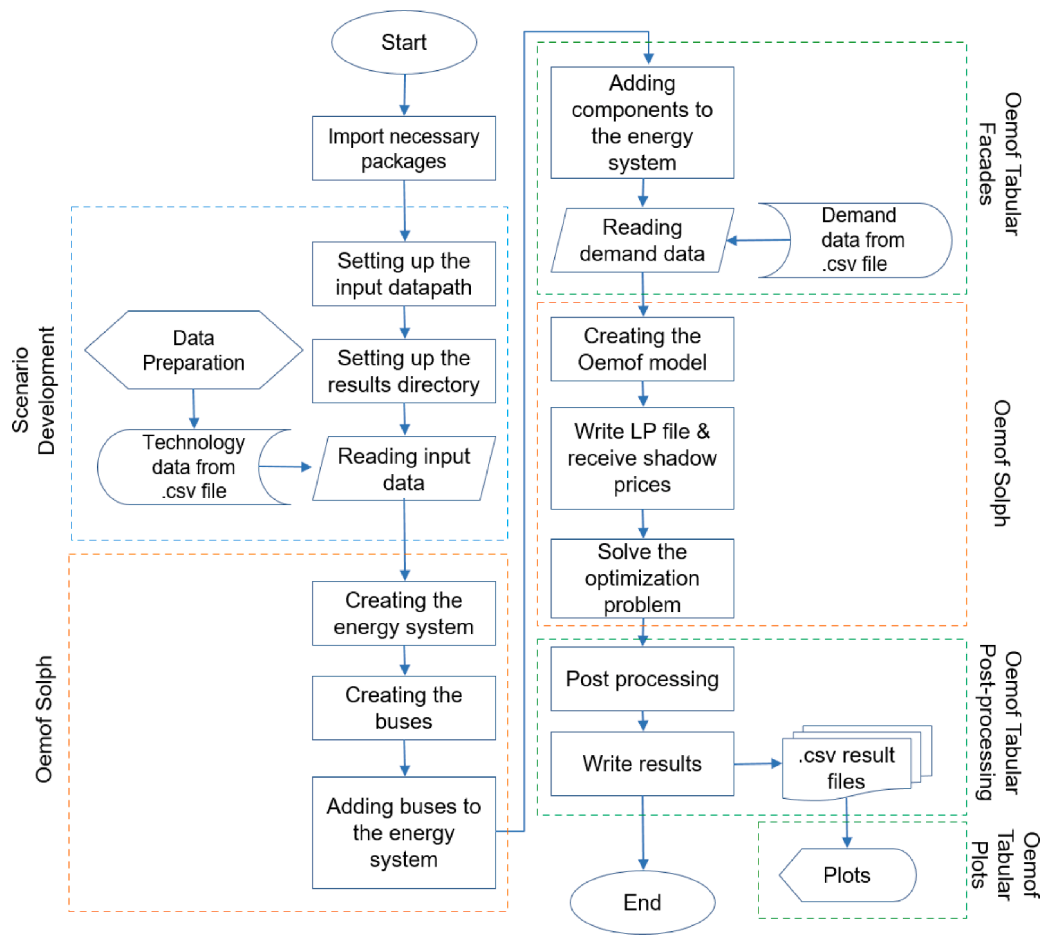

**Fig. 5.** OSeEM-DE Model development.

Python programming language is used for the model development in Fig. 5. After starting the necessary packages, input data and result directories are set up, and input data are read from external *.csv* files. Different scenarios are built with different datasets. The input datasets are pre-processed before importing them to the input *.csv* files. The energy system is created using Oemof Solph. Different components are added to the energy system using Oemof Tabular's *Facade* classes. Then the energy model is solved using Oemof Solph. After optimization, results are post-processed and written to several *.csv* files using *Post-processing* tool, which are later illustrated using *Plots* tool.

*4.1. Input data: hourly profiles and capacities*

The model optimizes investment and operation for a full year of historical hourly data, with 2011 chosen as the primary representative year because of its average wind conditions, high solar irradiation (representing the expected warm condition in 2050 due to climate change), slightly lower overall heating demand, and duration of maximum heating demand during the winter days [95].

*4.1.1. Electricity and heat load*
The hourly electricity load data based on the *ENTSO-e* statistical database [96] are obtained from the *Open Power System Data (OPSD)* project [97]. Hourly normalized time series are obtained from the electricity load data for both NDE and SDE energy systems based on population-based clustering. The total electricity demand for Germany in 2050, based on the representative year, is 532.77 TWh$_{el}$, of which NDE accounts for 96.6 TWh$_{el}$, and SDE accounts for 436.16 TWh$_{el}$.

The hourly heat load data based on the *When2Heat* project are obtained from the *OPSD* project [97–99]. Heat demand time series for space heat and hot water for Germany are obtained separately from the *When2Heat* profiles and are clustered and normalized for use as inputs of the model. The total heat demand for Germany in 2050, based on the representative year, is 648.53 TWh$_{th}$. Total space heating demand accounts for 533.6 TWh$_{th}$, of which NDE accounts for 96.75 TWh$_{th}$, and SDE accounts for 436.85 TWh$_{th}$. Total hot water demand accounts for 114.92 TWh$_{th}$, of which NDE accounts for 20.83 TWh$_{th}$, and SDE accounts for 94.08 TWh$_{th}$.

*4.1.2. Volatile generator*
Normalized onshore wind profiles based on *MERRA-2* datasets are obtained separately for NDE and SDE energy systems from the *Renewables Ninja* project [100]. Normalized offshore wind profiles based on *MERRA-2* datasets are also obtained from the *Renewables Ninja* project [100]. Similarly, normalized solar PV profiles based on *MERRA-2* datasets are obtained separately for NDE and SDE energy systems from the *Renewables Ninja* project [100]. Hydro inflow data for the ROR plants are obtained and normalized from the *Dispa-SET* project [101]. The current capacity and potential data are obtained or calculated from the *Agency for Renewable Energies* [102], *Deutsche WindGuard* [103], *LIMES-EU* project [104], *ZNES* [105], *PyPSA* [106–109], and *ANGUS II* project [110] databases. Table 1 shows the capacity and available potential data for the volatile generators in 2050.

The placement of volatile generators considers certain land limitations to account for competing land uses and minimum-distance regulations. For onshore wind, the maximum capacity density on the available area is 4 MW/km$^2$, as described in the *LIMES-EU* project [104].

**Table 1**
Capacity and potential for volatile generators in the German energy system in 2050.

| Energy System | | Onshore Wind [GW$_{el}$] | Offshore Wind [GW$_{el}$] | Solar PV [GW$_{el}$] | Hydro ROR [GW$_{el}$] |
|---|---|---|---|---|---|
| NDE | Capacity | 22.12 | 6.66 | 7.56 | 0.09 |
| | Available Potential | 49.88 | 76.94 | 55.77 | 0.06 |
| SDE | Capacity | 31.79 | – | 37.66 | 4.2 |
| | Available Potential | 118.81 | – | 188.07 | 1.1 |
| Germany | Total Capacity | 53.91 | 6.66 | 45.23 | 4.29 |
| | Total Available Potential | 168.69 | 76.94 | 243.84 | 1.16 |

The share of suitable areas available for RES is 30% and 5% for agricultural and forest areas, respectively. Public acceptance and nature reserves limit the share of the available onshore wind installation. For offshore wind, a maximum depth of 50 m is considered, and other factors such as 55 km maximum distance to shore, placement within the exclusive economic zone [104]. Besides, only 50% of the resultant area is considered with a maximum capacity density of 6 MW/km² to prevent turbine installations close to the mainland and allow a shipping corridor. For solar PV, the limitations include protecting nature reserves and restricted areas listed in the *Natura2000 Database*, and land use types according to the *CORINE Land Cover Database*, as described by Hörsch et al. [109]. The efficiency of the hydro ROR plant is 90% [110].

*4.1.3. CHP and heat pump*

The amount of available biomass is calculated from the *Hotmaps* project [111]. For a full year, the biomass potential for Germany is 634.33 PJ. For the current capacities, it is assumed that the existing biomass and biogas power plants are converted to CHP plants by 2050. Therefore, 4.74 GW CHP in NDE and 8.86 GW CHP in SDE energy systems are assumed to be installed and fully operational by 2050. It is also assumed that the CHP's electrical and thermal efficiencies are 45%, and the condensing efficiency is 50%. The current heat pump capacities are considered zero, and their investment potentials are not constrained. However, since heat pumps are P2H devices, their investment capacities depend upon limited available power from the potentially constrained volatile generators. The COP for ASHP and GSHP are 2.3 and 3.9, respectively [110].

*4.1.4. Storage*

The capacity and maximum potential data for storage investments are described in Table 2. Data for Li-ion and Redox flow batteries, H$_2$ storages, and ACAES are obtained from the *ANGUS II* project scenarios for Germany in 2050 [110]. PHS data are obtained from the *Dispa-SET* project [101]. For batteries, H$_2$ storage, and TES, it is assumed that 18% of the total potential is available for NDE, and the rest is available for SDE.

The energy storage potentials for the batteries, H$_2$ storage, and ACAES investment are subject to optimization. The storage potential for the existing ACAES plant in Huntorf is 0.58 GWh$_{el}$ [112]. PHS's inflow data are derived from the *Dispa-SET* project's scaled inflow dataset and Germany's current PHS capacity [101]. PHS's storage capacities are taken from the *Dispa-SET* project, which is 1.7 GWh$_{el}$ for NDE, and 717 GWh$_{el}$ for SDE [101]. PHS investment is excluded because of the limited expansion capacity. A loss rate of 1% is assumed for PHS. For TES, the considered loss rate is 1.4% for TES [106,110]. The loss rate indicates the relative loss of the storage content per time unit. The roundtrip efficiencies for Li-ion, Redox, H$_2$, PHS, ACAES, and TES are 92%, 80%, 46%, 75%, 73% and 81%, respectively [110]. The maximum state of charge capacity in terms of hours at full output capacities for Li-ion, Redox, H$_2$, PHS, ACAES, and TES are 6.5 h, 3.3 h, 168 h, 8 h, 7 h, and 72 h, respectively [110].

*4.1.5. Transmission line*

The transmission lines in between NDE and SDE energy systems (NDE-SDE Link) are modeled as transshipment capacities. The total transmission capacity is set and varied exogenously for different scenarios. According to the IEA data for Germany, a transmission and distribution line loss of 4% is considered [113].

*4.2. Input data: costs*

The investment capacity costs are based on annuity and fixed operation and maintenance (O&M) costs, and the marginal costs are based on variable O&M costs, the carrier costs, and the efficiencies. The carbon

**Table 2**
Capacity and maximum potential for storage investments.

| Energy System | | Li-ion [GW$_{el}$] | Redox [GW$_{el}$] | Hydrogen [GW$_{el}$] | ACAES [GW$_{el}$] | PHS [GW$_{el}$] | TES* [GW$_{th}$] |
|---|---|---|---|---|---|---|---|
| NDE | Capacity | 0 | 0 | 0 | 0.29 | 0.34 | 0 |
| | Available Potential | 2.82 | 0.17 | 1.82 | 3.43 | – | 1.8 |
| SDE | Capacity | 0 | 0 | 0 | – | 0.82 | 0 |
| | Available Potential | 12.83 | 0.76 | 8.28 | – | – | 8.2 |
| Germany | Capacity | 0 | 0 | 0 | 0.29 | 8.57 | 0 |
| | Available Potential | 15.65 | 0.93 | 10.1 | 3.43 | – | 10 |

* Own assumption.

**Table 3**
Parametric variation for different scenarios.

| Scenario | Electricity Demand [TWh$_{el}$] | Total Heat Demand [TWh$_{el}$] | Total Grid Capacity [GW$_{el}$] | Biomass Potential [PJ] | Electricity Storage Potential [GW$_{el}$] | Heat Storage Potential* [GW$_{th}$] |
|---|---|---|---|---|---|---|
| *Base* | 532.77 | 648.53 | 35 | 634.33 | Li-ion: 15.65<br>Redox: 0.93<br>ACAES: 1.71 | 7.5 |
| *Conservative* | 586.04 | 648.53 | 32 | 697.76 | Li-ion: 15.65 | 5 |
| *Progressive* | 479.49 | 583.68 | 38 | 570.89 | Li-ion: 15.65<br>Redox: 0.93<br>ACAES: 3.43<br>H$_2$: 10.1 | 10 |

* Own Assumption.

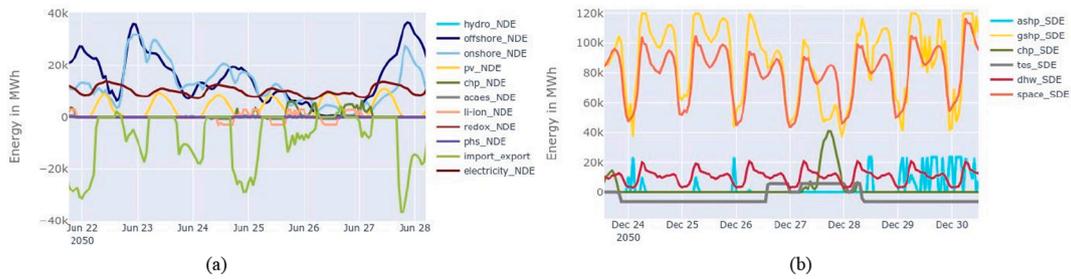

**Fig. 6.** Exemplary optimization results of the OSeEM-DE model for the electricity and heat buses in the base scenario in 2050 (a) hourly supply-demand variation of the electric bus for a week in June 2050 in NDE (b) hourly supply-demand variation of the heat bus for a week in December 2050 in SDE.

costs are not considered in the marginal costs since the system is 100% renewable. The required data for calculating the costs, such as capital expenditure, lifetime, the Weighted Average Cost of Capital (WACC), storage capacity cost, carrier cost, fixed and variable O&M costs, are taken from various references. Table A.1 (Appendix A) summarizes all input cost data for the OSeEM-DE Model.

## 5. Scenarios

Three scenarios, namely *Base*, *Conservative*, and *Progressive*, are developed and analyzed to answer the research questions in different conditions. The volatile generator inputs remain the same in all scenarios. The existing PHS and ACAES capacities also stay the same. The variation of the other parameters for different scenarios is shown in Table 3.

Regarding weather data and demands in 2050, the model considers 2011 as the representative weather year in the base scenario, according to Brown et al. [95]. The model also investigates the change in future electricity and heat demands considering that the demand will go high if there is a higher population or low if due to increased use of efficient, innovative, and demand-responsive technologies.

### 5.1. Base scenario

The base scenario assumes that electricity and heating demands remain the same as the representative year (2011) in 2050. For storage, Li-ion batteries, PHS, and ACAES (NDE) are considered. The full biomass potential from forest residues, livestock effluents, and agricultural residues in Germany is used as CHP input. For the electrical storage, Li-ion and Redox batteries up to their maximum potentials are used. Besides, half of the total ACAES potential is considered for investment.

### 5.2. Conservative scenario

The conservative scenario assumes that the electricity demand will increase by 10% than the base scenario in 2050. The heating demand is kept the same. For the grid transmission from NDE to SDE, a reduced capacity is considered. An additional 10% of biomass import is considered on top of Germany's full potential. For storage, ACAES and Redox investment possibilities are omitted, and only Li-ion battery investment

**Table 5**
Change in energy generation with reference to the base scenario.

| Generation Technologies | Change from the Base Scenario [%] | |
|---|---|---|
| | Conservative Scenario | Progressive Scenario |
| Onshore Wind | 0% | 0% |
| Offshore Wind | +36% | −46% |
| Solar PV | +17% | −10% |
| CHP (Electricity) | +10% | −10% |
| Hydro ROR | 0% | 0% |
| Heat Pumps (GSHP, ASHP) | −1% | −10% |
| CHP (Heat) | +10% | −13% |

is considered.

### 5.3. Progressive scenario

The progressive scenario assumes that both the electricity and heat demand will reduce by 10% than in the base scenario in 2050. For the grid transmission from NDE to SDE, increased maximum capacity is considered. The maximum usage of biomass is reduced by 10%. All electrical and heat storage are optimized for investment up to their full potentials.

## 6. Results & discussions

### 6.1. Sufficiency of the energy system

The model runs reached feasible solutions for all three scenarios. Fig. 6 shows examples of hourly energy supply-demand variation after cost-optimization in Germany's electricity and heat buses for the base scenario. Fig. 6 (a) shows the electricity supply using the volatile generators, CHP, and electrical storages to satisfy the electricity demand in NDE during a week in June 2050. Fig. 6 (b) shows the heat supply using heat pumps, CHP, and TES to satisfy the space heat and domestic hot water demands in SDE during a week in December 2050. Detailed scenario-wise and percentage-wise optimization results for NDE and SDE is available as supplementary material of the paper.

Table 4 summarizes the results for the three scenarios in terms of energy generation.

**Table 4**
Energy generation for different scenarios from the OSeEM-DE Model.

| Scenario | | Onshore Wind [TWh_el] | Offshore Wind [TWh_el] | Solar PV [TWh_el] | CHP (Electricity) [TWh_el] | Hydro ROR [TWh_el] | GSHP [TWh_th] | ASHP [TWh_th] | CHP (Heat) [TWh_th] | NDE to SDE [TWh_el] | SDE to NDE [TWh_el] |
|---|---|---|---|---|---|---|---|---|---|---|---|
| *Base* | NDE | 155.19 | 188.43 | 17.4 | 39.19 | 0.4 | 66.86 | 16.11 | 39.19 | 122.22 | 3.3 |
| | SDE | 197.98 | – | 269.38 | 61.71 | 14.56 | 460.7 | 13.47 | 61.1 | | |
| *Conservative* | NDE | 155.19 | 257.98 | 67.97 | 43.11 | 0.4 | 65.04 | 18.5 | 43.11 | 143.16 | 0.04 |
| | SDE | 197.98 | – | 269.38 | 67.83 | 14.56 | 457.73 | 8.2 | 67.7 | | |
| *Progressive* | NDE | 155.19 | 101.17 | 8.11 | 35.27 | 0.4 | 58.74 | 13.2 | 35.27 | 94.56 | 6.26 |
| | SDE | 197.98 | – | 250.22 | 55.92 | 14.56 | 409.28 | 22.53 | 51.6 | | |

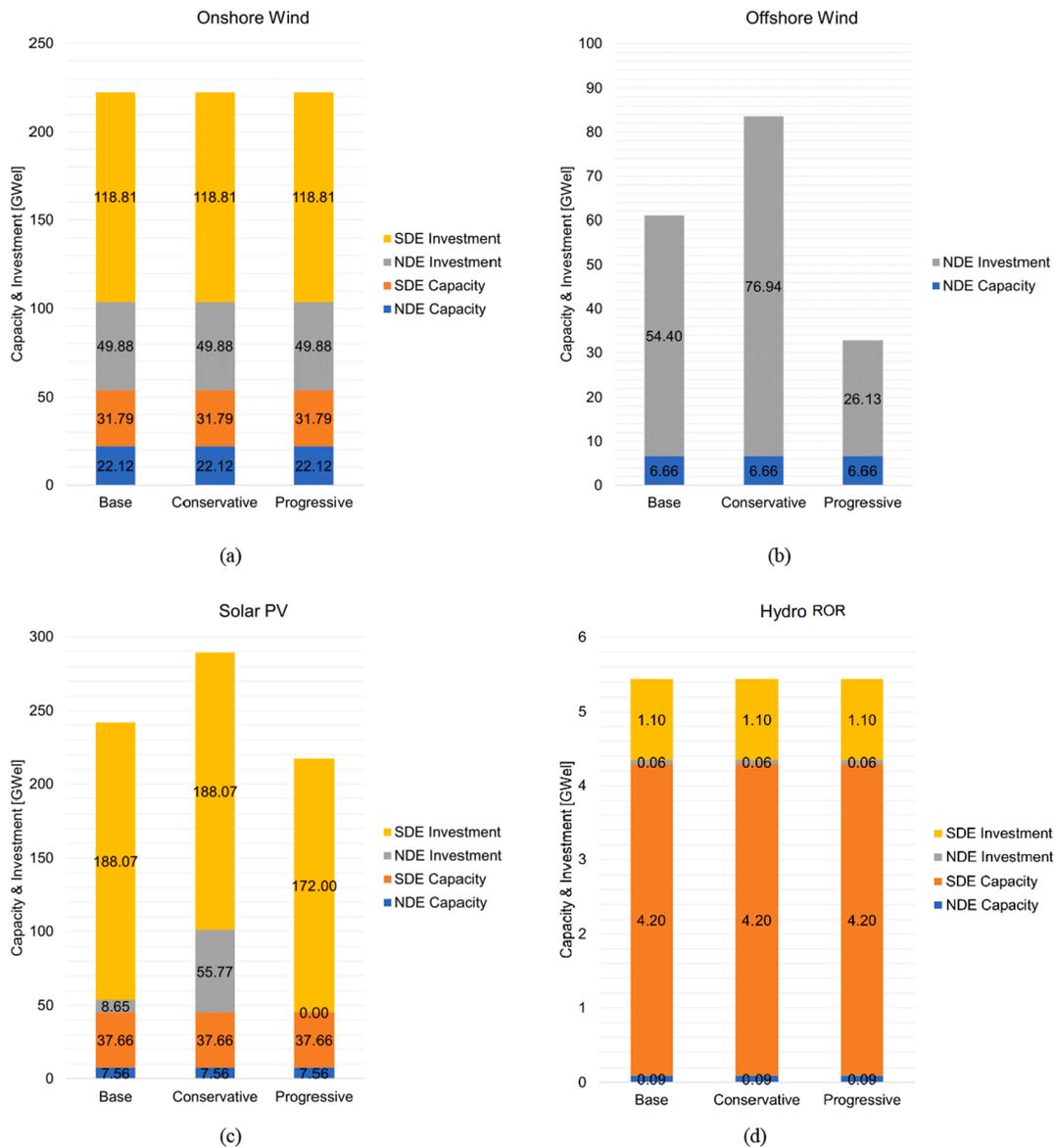

**Fig. 7.** Scenario-wise comparison of installed capacity and required investments for volatile generators in Germany (a) onshore wind (b) offshore wind (c) PV (d) hydro ROR plants.

Onshore wind and hydro ROR remain the same in all three scenarios. The change from the base scenario for the other generation technologies is summarized in Table 5.

Therefore, the model results show that Germany's renewable resources are sufficient to meet its electricity and building heat demands. The model suggests maximum usage of onshore wind and hydro ROR, and the rest of the demands are met using offshore wind, solar PV, CHP, and heat pumps. The decrease of 46% in the progressive scenario indicates that reducing electricity and heating demands can heavily affect Germany's offshore wind investment. An increase in electricity export from NDE to SDE is observed in the conservative scenario, indicating the matching of SDE demand using NDE's resources. The increase in biomass usage in the conservative scenario indicates the necessity of biomass import from other countries if the electricity demand is higher. The opposite situation can be observed in the progressive scenario when the electricity and heat demands are lower than the base scenario. Also, almost all the storages, except Hydrogen, are used to their maximum capacities in the progressive scenario. Two critical observations from the model run are:

1. Without the biomass-based CHPs, and there was no feasible solution for a power-building heat coupled system. That being the case, it is evident that Germany needs to use other alternative heating technologies besides heat pumps, combined with various heat storage options, to reduce its dependency on only biomass for meeting the heat demand.
2. While the model prefers ACAES and batteries, it does not choose Hydrogen for these three scenarios. The results will alter for a system with industrial process heating demand and inexpensive Hydrogen.

### 6.2. Capacities and investments

#### 6.2.1. Volatile generator

*6.2.1.1. Volatile generator capacity and investment.* Fig. 7 compares the existing capacities and required investments of the volatile generators. For onshore wind plants illustrated by Fig. 7 (a), in addition to the existing 53.91 $GW_{el}$, investment in 168.69 $GW_{el}$ is necessary for all three scenarios for Germany, of which 49.88 $GW_{el}$ is in the NDE, and 118.81 $GW_{el}$ is in the SDE energy system. Therefore, the model suggests maximum utilization of the available potential of all onshore wind energy in Germany. In the case of offshore wind plants shown by Fig. 7 (b), on top of the existing 6.66 GW$_{el}$, the base scenario suggests installing 54.4 GW$_{el}$, the conservative scenario recommends installing 76.94 GW$_{el}$ and the progressive scenario suggests installing 26.13 GW$_{el}$ capacities.

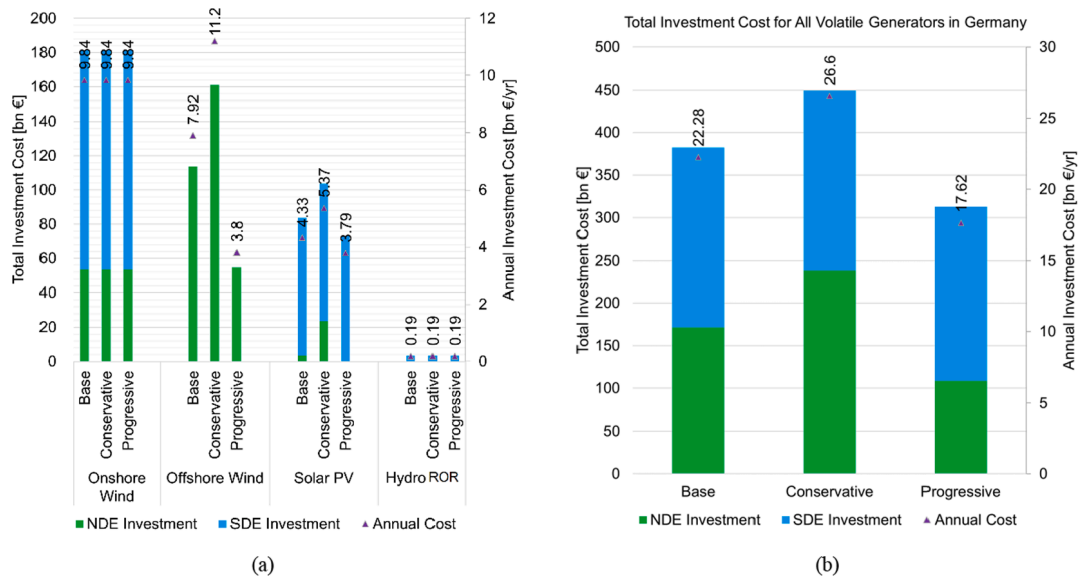

**Fig. 8.** Scenario-wise comparison of investment costs for volatile generators in Germany (a) technology-wise comparison (b) total investment cost for all volatile generators. The primary vertical axis (left) shows the total investment cost for the whole time-horizon in billion euros (shown as stacked columns), and the secondary vertical axis (right) shows the annual investment cost in billion euros/year (shown in markers).

For solar PV, as shown in Fig. 7 (c), in addition to the existing 45.23 GW$_{el}$ capacity, the base and conservative scenarios suggest installing 188.07 GW$_{el}$ in the SDE energy system. Contrarily, the progressive scenario in SDE suggests a reduced solar PV installation of 172 GW$_{el}$. In Northern Germany, the model suggests installing 8.65 GW$_{el}$ in the base scenario and 55.77 GW$_{el}$ in the conservative scenario. Interestingly, the progressive scenario suggests that no additional solar PV installation is necessary for NDE. In the case of Hydro ROR plants shown in Fig. 7 (d), both the NDE and SDE systems suggest capacity increment to the maximum available potentials, i.e., 0.06 GW$_{el}$ in NDE and 1.1 GW$_{el}$ in SDE, on top of the existing 4.29 GW$_{el}$ in Germany.

*6.2.1.2. Volatile generator investment cost.* Fig. 8 shows the investment costs of the volatile generators. Fig. 8 (a) shows that the annual investment cost remains the same for onshore wind (9.84 bn €/yr) and hydro ROR plants (0.19 bn €/yr). The annual investment is 7.92 bn €/yr for offshore wind, which increases in the conservative scenario (11.2 bn €/yr) and decreases in the progressive scenario (3.8 bn €/yr). For solar PV, the base scenario investment of 4.33 bn €/yr goes high in the conservative scenario (5.37 bn €/yr) and goes low in the progressive scenario (3.79 bn €/yr).

Fig. 8 (b) compares the total investment cost for all volatile generators. The annual investment cost for all the volatile generators is 22.28 bn €/yr, which increases by 19% in the conservative scenario (26.6 bn €/yr) and decreases by 21% in the progressive scenario (17.62 bn €/yr).

*6.2.2. CHP and heat pump*

*6.2.2.1. CHP and heat pump capacity and investment.* Fig. 9 compares the existing capacities and required investments for the heat generators. For CHP shown by Fig. 9 (a), in addition to the existing 13.6 GW, investment in 68.61 GW, 73.62 GW, and 58.26 GW, additional capacities are necessary for the base, conservative, and progressive scenarios, respectively. Fig. 9 (b) shows a similar pattern for heat pumps (GSHP and ASHP) in Fig. 9 (b), where the base scenario requires 173.41 $GW_{th}$ investment, the conservative scenario requires an increased 181.86 $GW_{th}$ investment, and the progressive scenario requires a decreased 154.93 $GW_{th}$ investment in Germany.

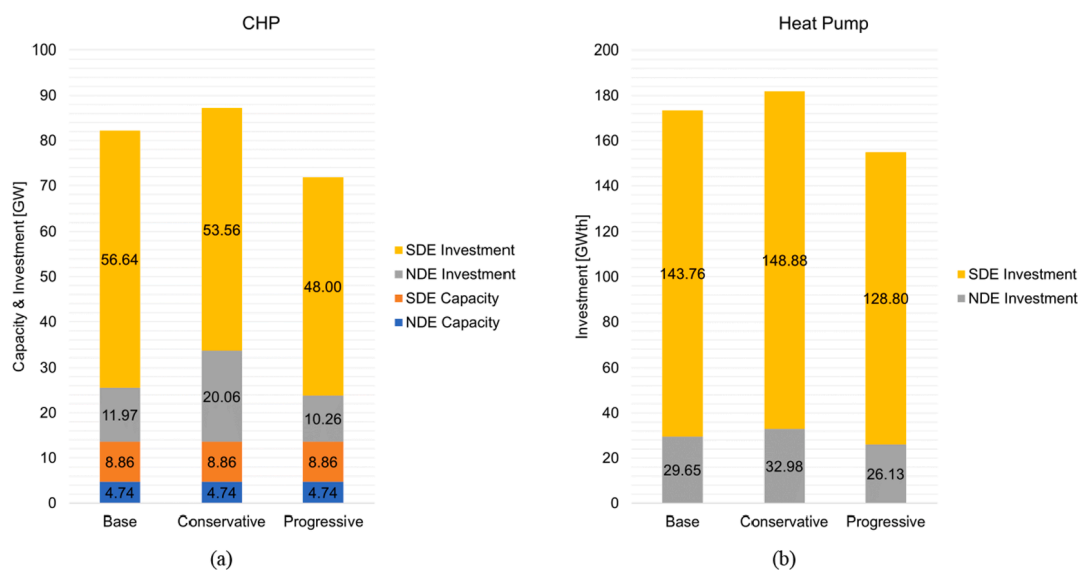

**Fig. 9.** Scenario-wise comparison of installed capacity and required investments for CHP and Heat Pump in Germany (a) CHP (b) Heat Pump (GSHP and ASHP).

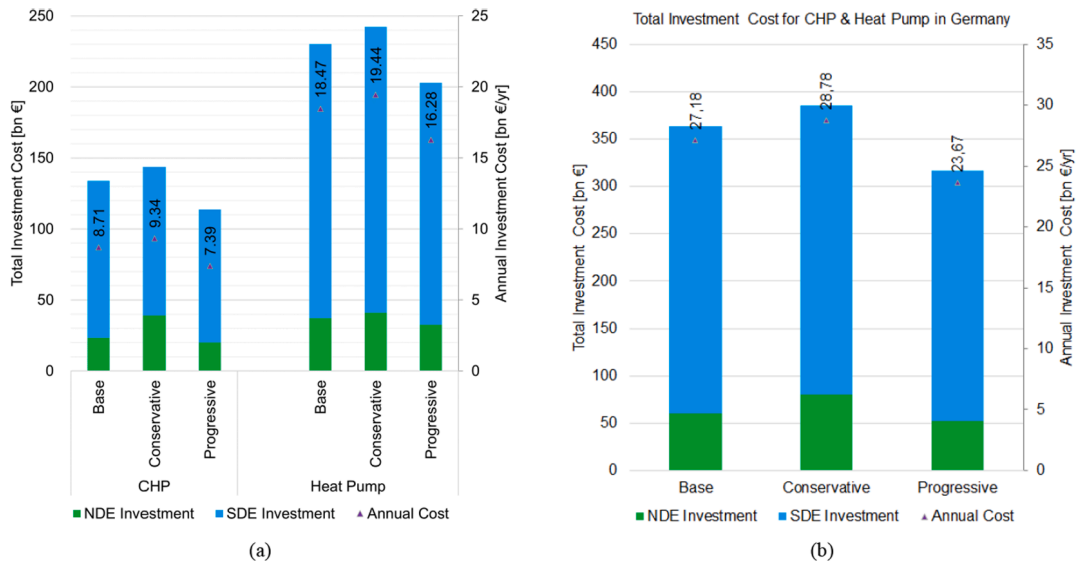

**Fig. 10.** Scenario-wise comparison of investment costs for CHP and heat pump in Germany (a) technology-wise comparison (b) total investment cost for all heat generators (CHPs & heat pumps).

Although the heat demand did not change in the conservative scenario, heat generator installations increases because of the following reasons:

1. Since volatile generators are already installed up to their maximum potential, increased CHP installations are required to meet the additional electricity demand;
2. Reduced grid transfer capacity minimizes the power exchange scope between the grids and hence lessens the scope of P2H conversion with power from the other energy system. This circumstance enforces the model to install additional local heat pumps in individual energy systems;
3. Heat generation in this energy system depends on both electrical (because of P2H) and heat storage. Since electrical and heat storages are reduced in the conservative scenario, supply-demand balancing requires additional heat generators.

This additional heat generator in the conservative scenario also results in surplus heat in both energy systems at different hours over the year. Nevertheless, this additional heat investment problem is minimized in the progressive scenario where we have reduced electricity and heat demands, increased grid transfer capacity, and increased electrical and heat storage.

*6.2.2.2. CHP and heat pump investment cost.* Fig. 10 shows the technology-wise investment costs of the CHPs and heat pumps for NDE and SDE and the total costs. As shown in Fig. 10 (a), the annual investment cost for CHPs in the base scenario is 8.71 bn €/yr, which increases to 9.34 bn €/yr in the conservative and decreases to 7.39 bn €/yr in the progressive scenario. Similarly, for heat pumps, the base scenario investment of 18.47 bn €/yr increases to 19.44 bn €/yr in the conservative scenario and decreases to 16.28 bn €/yr in the progressive scenario.

Fig. 10 (b) compares the total investment cost for all CHPs and heat pumps. The annual investment cost for all the heat generators is 27.18 bn €/yr, which increases by 5% in the conservative scenario (28.78 bn €/yr), and decreases by 13% in the progressive scenario (23.67 bn €/yr).

*6.2.3. Storage*

*6.2.3.1. Storage capacity and investment.* Fig. 11 compares the investment for all types of storage. The model does not suggest any investment in $H_2$ storage because of the high cost. Fig. 11 (a) and Fig. 11 (b) shows the required investment capacities of Li-ion and Redox batteries. Both batteries are used up to their maximum given potential whenever used as an input. According to Fig. 11 (a), for a total of 15.65 $GW_{el}$ Li-ion batteries, the total optimized energy capacity is 102 $GWh_{el}$ (NDE = 18.32 $GWh_{el}$, SDE = 83.42 $GWh_{el}$). As shown in Fig. 11 (b), for 0.93 $GW_{el}$ Redox batteries in Germany, the total optimized energy capacity is 3 $GWh_{el}$ (NDE = 0.56 $GWh_{el}$, SDE = 2.52 $GWh_{el}$).

However, the ACAES are used partially with an optimized power of 0.15 $GW_{el}$ and optimized energy storage of 2.5 $GWh_{el}$ in the base scenario, and an increased optimized capacity of 0.51 $GW_{el}$ and optimized energy storage of 5 $GWh_{el}$ in the progressive scenario, as shown in Fig. 11 (c). Fig. 11 (d) shows the usage of TES capacities up to their maximum exogenous potential, with the total optimized energy storage capacity varying in between 360 $GWh_{th}$ (conservative) and 720 $GWh_{th}$ (progressive).

*6.2.3.2. Storage investment cost.* Fig. 12 shows the investment costs of the electrical and heat storage and the total costs. As shown in Fig. 12 (a), Li-ion batteries' annual investment cost is 1.57 bn €/yr. For Redox batteries, the base scenario's annual investment cost and the progressive scenario are 0.05 bn €/yr. In the case of ACAES, the annual investment cost is 0.01 bn €/yr in the base scenario and 0.03 bn €/yr in the progressive scenario. The annual investment cost of heat storage depends mainly upon the volume of the storage capacity, which is 1.66 bn €/yr in the base scenario for 540 $GWh_{th}$. As the storage capacity goes down to 360 $GWh_{th}$ in the conservative scenario, the yearly investment reduces to 1.1 bn €/yr. In the progressive scenario, the annual investment again increases to 2.21 bn €/yr since the total storage capacity increases to 720 $GWh_{th}$.

Fig. 12 (b) shows the total investment costs for all storages. The annual investment for all storages is 3.29 bn €/yr in the base scenario, which decreases by 18% to 2.67 bn €/yr in the conservative scenario with reduced storage provisions and increases by 17% to 3.86 bn €/yr in the progressive scenario with high storage provisions.

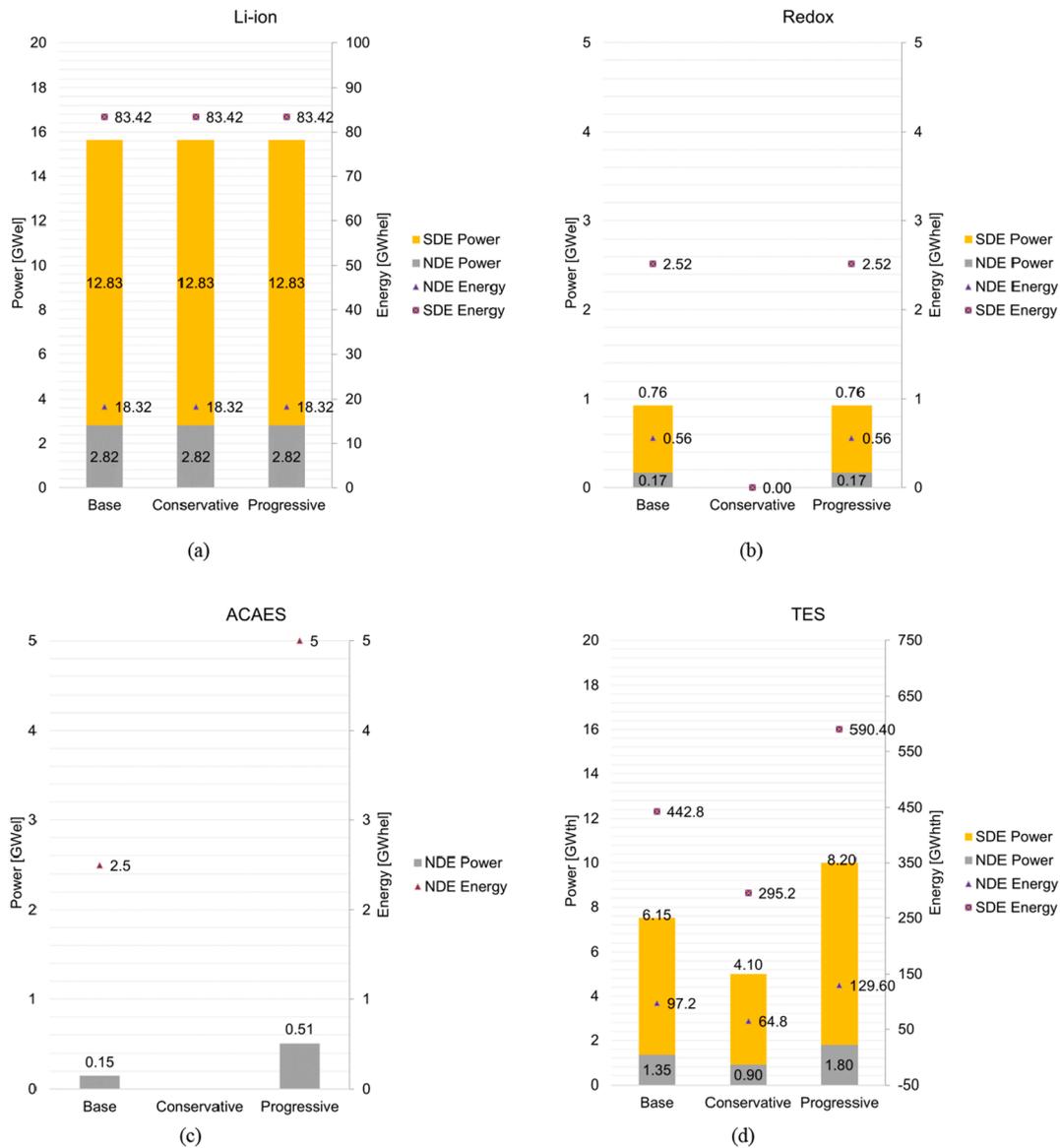

**Fig. 11.** Scenario-wise comparison of required power and energy storage investments for electrical and heat storages in Germany (a) Li-ion (b) Redox (c) ACAES (d) TES. The primary vertical axis (left) shows the power capacity in $GW_{el}/GW_{th}$ (shown as stacked columns), and the secondary vertical axis (right) shows the energy storage capacity $GWh_{el}/GWh_{th}$ (shown in markers).

### 6.3. Energy mix and flexibility aspects

#### 6.3.1. Energy mix

Fig. 13 compares the energy mix results. Fig. 13 (a) shows the composition of the electricity generation mix. The total electricity generation in the base scenario is 944 $TWh_{el}$, increasing to 1074 $TWh_{el}$ in the conservative scenario with excess demand, less storage, reduced grid capacity; and decreased to 818 $TWh_{el}$ in the progressive scenario with reduced demand, more storage, and increased grid capacity. The energy mix also shows that, while all three scenarios need maximum offshore wind and hydro ROR, additional solar PV, offshore wind, and CHP can supply the rest of the demand. The base scenario mix comprises 37% onshore wind, 19% offshore wind, 30% solar PV, 10% CHP, and 0.01% hydro ROR plants. The progressive scenario shows similar results: 43% onshore wind, 12% offshore wind, 31% solar PV, 11% CHP, and 0.01% hydro ROR. These results are compared and cross-validated with a study from Fraunhofer Institute for Solar Energy System (ISE) [114], which showed that an 85% renewable-based system in Germany producing 800 $TWh_{el}$ electricity comprises 47% onshore wind, 16% offshore wind, and 22% solar PV. According to the OSeEM-DE results, CHPs can replace the remaining 15% fossil-fuel-based generation using Germany's available biomass potential. Such a system also needs high electrification of the heating sector. Fig. 13 (b) shows that with less demand, more heat storage capacity, and more grid transfer capacity, the progressive scenario yields a reduced heat generation of 590 $TWh_{th}$.

The heat generation results also show that heat pumps dominate the heat generation (around 80%), and biomass-CHPs cover the rest of the demand. In heat pumps, the comparatively more efficient but expensive GSHPs are preferable over less efficient and cheaper ASHPs. Though the model considers only CHPs and heat pumps, other renewable applications such as electric boiler, solar heating, biofuel heating, Hydrogen heating, and geothermal heating should be investigated to produce heat, primarily process heating in industries.

Excess generation from renewables is a challenge that needs to be solved using a combination of plausible solutions. The OSeEM-DE results show that the least amount of excess generation occurs in the progressive scenario, indicating the apparent solution of using more storage. Also, the developed model is currently an island model with no interconnection between neighboring countries. A German energy system and its neighboring countries can reduce the curtailment problem with reduced optimized generation and storage capacities. Furthermore, demand response activities and the transport sector's inclusion with electric vehicles are also needed to reduce the curtailment or utilize the excess generation. The Levelized Cost of Electricity (LCOE) is calculated from the optimization results according to (23) (See Section 3.9). The LCOE values from the OSeEM-DE model are compared to a study from Fraunhofer ISE, which illustrates the current LCOE values (2018), and the learning curve-based predicted LCOE values (2035) in Germany [115]. We can see from Table 6 that the LCOE results from the OSeEM-DE model for onshore and offshore wind and solar PV technologies are very close to the lower range of LCOE values in 2035 as forecasted in the Fraunhofer ISE study. While the Fraunhofer ISE study considers biogas for electricity generation, the OSeEM-DE considers biomass for electricity and heat generation, resulting in higher LCOE values.

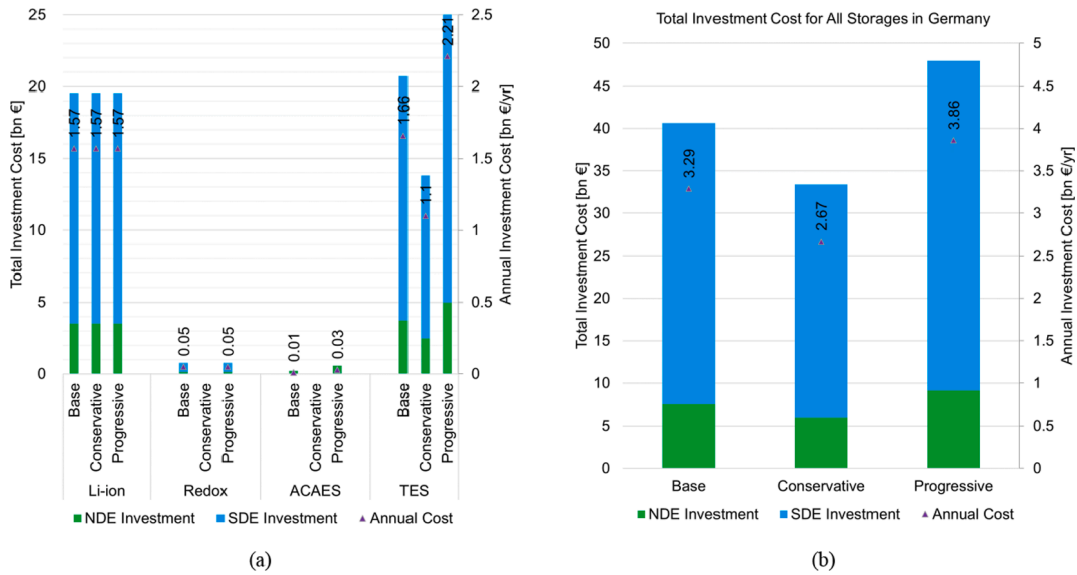

**Fig. 12.** Scenario-wise comparison of investment costs for electrical and heat storages in Germany (a) technology-wise comparison (b) total investment cost for all storages.

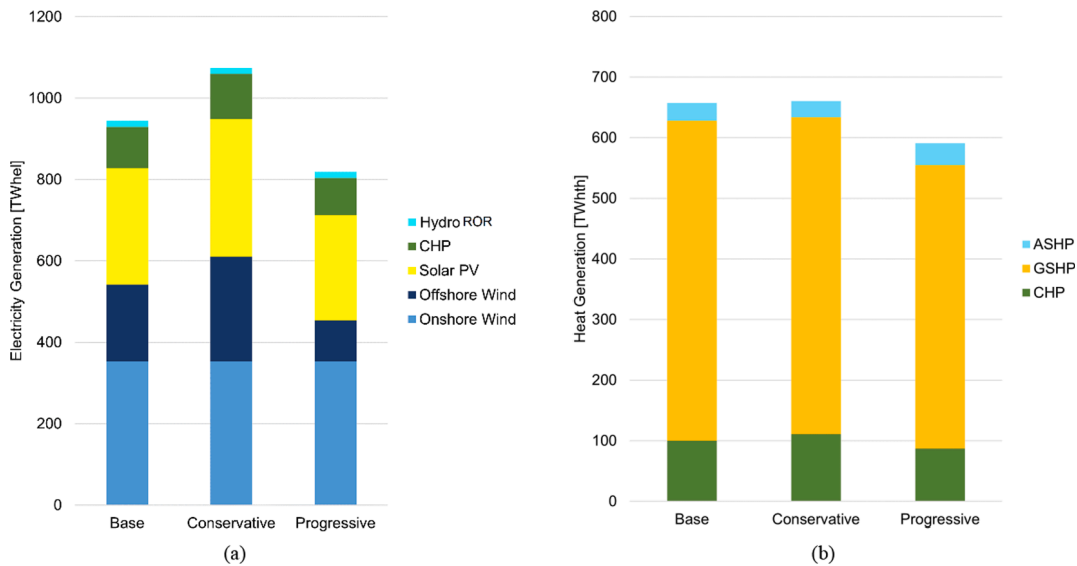

**Fig. 13.** Scenario-wise comparison for energy generation (a) electricity generation (b) heat generation.

**Table 6**
LCOE comparison of Fraunhofer ISE study and OSeEM-DE model results.

| Technology | Levelized Cost of Electricity [€ cent/kWh] | | | | |
|---|---|---|---|---|---|
| | Fraunhofer ISE Study (2018) [115] | Fraunhofer ISE Study (2035) [115] | OSeEM-DE Model (2050) | | |
| | | | Base | Conservative | Progressive |
| Onshore Wind | 3.99–8.23 | 3.49–7.09 | 4.99 | 4.99 | 4.99 |
| Offshore Wind | 7.49–13.79 | 5.67–10.07 | 6.79 | 6.93 | 6.34 |
| Solar PV | 3.71–11.54 | 2.41–4.70 | 3.61 | 3.73 | 3.56 |
| Biogas/Biomass | 10.14–14.74 | 10.14–14.74 | 20.26 | 19.76 | 19.47 |

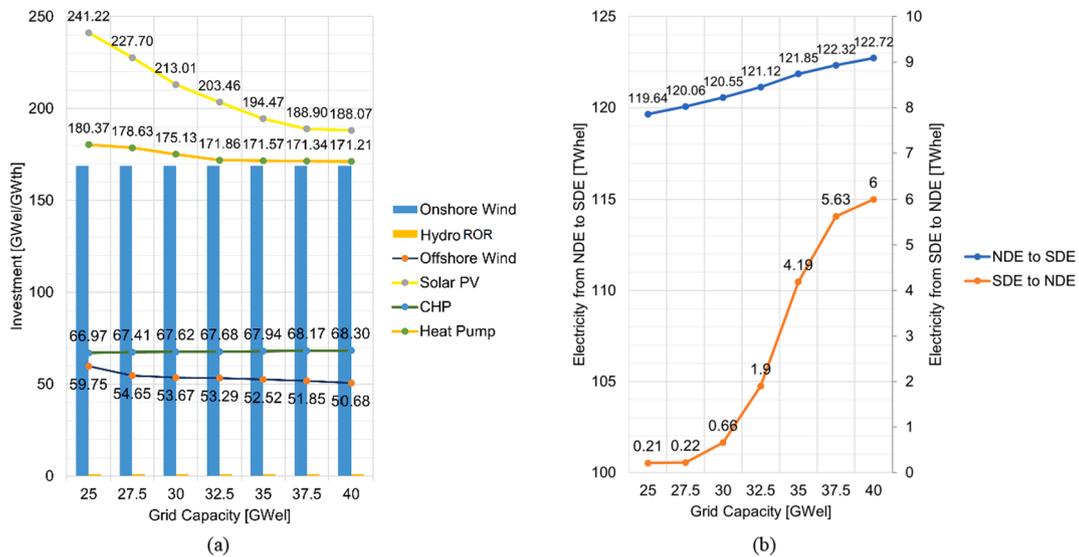

**Fig. 14.** Effect of grid capacity expansion between Northern and Southern Germany (a) investment in electricity and heat generators. The varying values of offshore wind, solar PV, CHP, and heat pumps are shown using lines, and the steady onshore wind and hydro ROR capacities are shown using columns (b) total annual electrical energy transfer between NDE and SDE. Two different vertical axes are used for showing the exchange, the primary vertical one (left) for NDE to SDE, and the secondary vertical one (right) for SDE to NDE.

*6.3.2. Flexibility aspects*

*6.3.2.1. Grid expansion.* This study's primary assumption is the growth of grid connection in 2050, resulting in a significant increase of power exchange between northern and southern Germany. The results suggest a large amount of grid exchange from NDE to SDE. For example, 28% of the SDE electrical demand comes from NDE (122.22 TWh$_{el}$ vs. 436.16 TWh$_{el}$) in the base scenario. Similarly, in the conservative scenario, around 29% of the SDE electrical demand comes from NDE (143.17 TWh$_{el}$ vs. 479.78 TWh$_{el}$), and in the progressive scenario, around 24% of the SDE demand comes from NDE (94.57 TWh$_{el}$ vs. 392.55 TWh$_{el}$). A sensitivity analysis is conducted to inspect the effect of grid expansion. For this analysis, the base scenario is modified where all the storage capacities are optimized to their maximum potentials, and the grid capacity varies between 25 GW$_{el}$ and 40 GW$_{el}$. Fig. 14 (a) shows that with the grid expansion from 25 GW$_{el}$ to 40 GW$_{el}$, the offshore wind investment reduces by 15% from 59.75 GW$_{el}$ to 50.68 GW$_{el}$. Solar PV investment reduces by 22% from 241.22 GW$_{el}$ to 188.07 GW$_{el}$. The onshore wind and hydro ROR investments remain the same. Heat pump investment decreases by 5% (171.37 GW$_{th}$ vs. 171.21 GW$_{th}$). While grid expansion decreases offshore wind plant capacities, relatively less expensive biomass-based CHPs replace a share of the curtailed offshore generation. Therefore, CHP installation increases by 2% (66.97 GW vs. 68.30 GW) with the grid expansion from 25 GW$_{el}$ to 40 GW$_{el}$.

Fig. 14 (b) shows the increase in the total electricity transfer between NDE and SDE for increasing grid capacities. For the grid expansion from 25 GW$_{el}$ to 40 GW$_{el}$, an increase of 2.5% (119.64 TWh$_{el}$ vs. 122.72 TWh$_{el}$) from NDE to SDE is observed. On the other hand, from SDE to NDE, though the transfer amount is much less compared with NDE to SDE, a sharp increment rate can be observed (0.21 TWh$_{el}$ vs. 6 TWh$_{el}$) when grid capacity expands.

The results for power and energy capacities for all the storage remained constant for all grid capacities, which indicated the maximum usage of the exogenously provided potentials. Therefore, the impact on the storage capacity for the grid expansion could not be measured using this sensitivity analysis. However, the grid expansion facilitates a smoother balance between the two systems, resulting in increased electricity exchange, thus requiring less investment in offshore wind, solar PV, and heat pump installations. Therefore, the expansion of the electrical grid between Northern and Southern Germany should be considered as a promising option for supporting a 100% renewable-based sector-coupled system for Germany. However, the cost of grid expansion vs. investment in generation facilities, which has not been conducted in this research, must be investigated to reach an optimum solution and draw a conclusion.

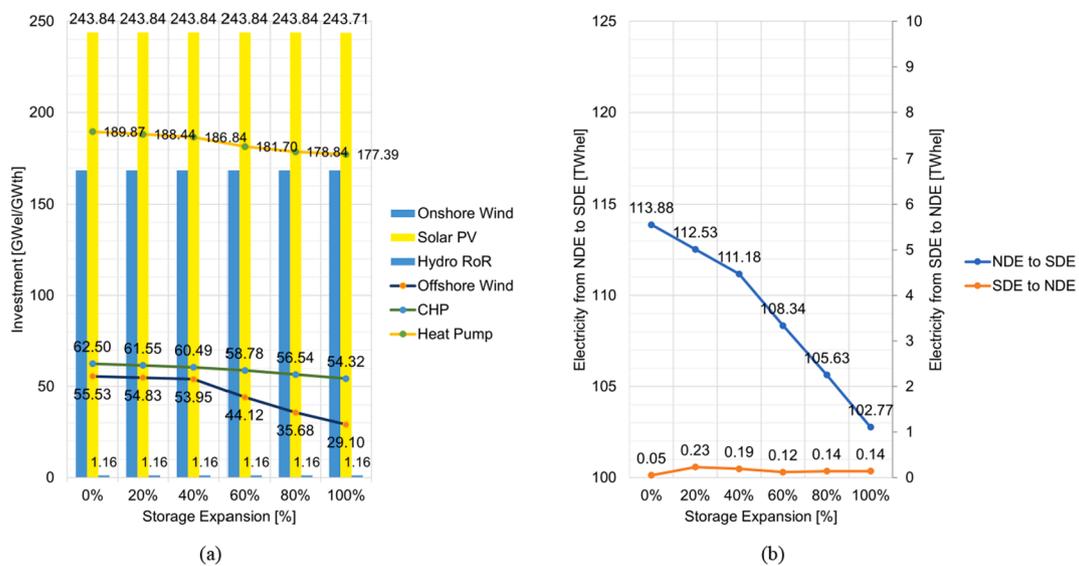

**Fig. 15.** Effect of storage expansion from current capacity (0%) to double capacity (100%) (a) investment in electricity and heat generators. The varying values of offshore wind, CHP and heat pumps are shown using lines, and the steady onshore wind, solar PV and hydro ROR capacities are shown using columns (b) total annual electrical energy transfer between NDE and SDE.

*6.3.2.2. Storage and dispatchable load.* Electrical storages and dispatchable loads (i.e., heat pumps) with heat storage tend towards flexibility and interdependence in all the scenarios. The existing hydro ROR plants and PHS help the system to balance both the short-term and long-term. Also, the biomass-fed CHPs are used as dispatchable generation resources that serve as a backup to counter the volatile generation's variability from solar PV and wind. Both Li-ion and Redox batteries act as critical storage technologies to be utilized for shorter periods. On the other hand, ACAES shows promising prospects to aid the PHS for long-term seasonal storage. The large-scale investment of heat pumps confirm the findings of Hedegaard and Münster, i.e., the individual heat pumps can have a positive contribution towards large scale wind power investments to reduce the system cost and pressure on the limited biomass potential [116]. The dispatchable heat pumps can use surplus power from the variable renewable generation, which supplements other heat generation sources and offers flexible operations. A sensitivity analysis is conducted to inspect the effect of storage expansion. For this analysis, the grid capacity is fixed at 20 GW$_{el}$, and the batteries (Li-ion and Redox), H$_2$ storage, and TES are gradually doubled from their current maximum exogenous potential. The ACAES and PHS capacities are not expanded since spatial constraints limit their maximum potentials. As shown in Fig. 15 (a), the onshore wind, solar PV, and hydro ROR capacities do not change with the increase in storage capacities, but the offshore investment drops by 47% (55.53 GWel vs. 29.1 GWel) when the storages are doubled. Similarly, both CHP and heat pump investment decreases by 13% (62.5 GW vs. 54.32 GW) and 6% (189.87 GW$_{th}$ vs. 177.39 GW$_{th}$), respectively. Hence, increased storage options offer additional flexibility to curtail peak/reserve capacities for the energy system.

On the other hand, electricity exchange decreases from NDE to SDE by more than 9% (113.88 GW$_{el}$ vs. 102.77 GW$_{el}$) when the storage capacity is doubled, as shown in Fig. 15 (b). Contrarily, the electricity transmission from SDE to NDE is almost a straight line, with a much lower value. The decrease of investment in offshore wind, CHP, and heat pumps and the reduced energy transfer from NDE to SDE indicate that storage expansion in local energy systems can be another viable flexibility option for reducing investment in generation capacities and grid expansion.

Considering the limited reserve capacity of cobalt and vanadium, expanding the battery capacities is a matter of further investigation. The sensitivity analysis for storage expansion also reveals the usage of H$_2$ storages in three instances (0%, 20%, and 40%) when the grid capacity was comparatively low (20 GW$_{el}$), and the other storage options were not sufficient. This result indicates the possibility of H$_2$ storage as another plausible alternative, especially for local energy systems, when other resources for batteries are not adequate and when the grid expansion is limited. In addition to H$_2$ storage, Norwegian large hydro capacities are another promising storage option for the future German energy system. This requires additional investment in interconnection capacities between Norway and Germany and is subject to policy discussions.

## 7. Summary

### 7.1. Summary of the analysis

The analysis shows that a 100% renewable energy system for both power and building heat sectors are feasible for moderate to extreme scenario considerations. For onshore wind and hydro ROR investments, maximum potentials should be utilized to meet the demand. In contrast, in the case of PV and offshore wind turbines, the investment capacities depend upon variation in electricity and heat demand, transmission grid expansion, available biomass potential, and storage provisions. The energy mix of such a system is composed of all the possible volatile generator resources, biomass resources, and the already existing renewable capacities. The model chooses PHS, ACAES, and batteries over relatively expensive H$_2$. The results will be different when the industrial heating sector is coupled, and large-scale electrolyzers produce inexpensive H$_2$.

In volatile generator investments, the scenario analysis shows that a fixed investment of 168.69 GW$_{el}$ onshore wind and 1.16 GW$_{el}$ hydro ROR plants are necessary, along with varying investments of 26.13–76.94 GW$_{el}$ offshore wind and 172–243.84 GW$_{el}$ solar PV capacities for the three developed scenarios. Onshore wind plants require an annual investment of 9.84 bn €/yr, and hydro ROR plants require a yearly investment of 0.19 bn €/yr. Offshore wind and solar PV requires varying investments in different scenarios, 3.8–11.2 bn €/yr for offshore wind and 3.79–5.37 bn €/yr for solar PV. The total cost for the volatile generators in Germany varies between 17.62 bn €/yr and 26.6 bn €/yr, which sums up to 312–450 bn € over the lifetime.

Heat generator capacities vary for CHPs between 58.26 GW$_{th}$ and 73.62 GW$_{th}$, while the heat pump acts as a more preferred alternative heating option for buildings, requiring 154.93–181.86 GW$_{th}$ installations for Germany. The cost varies accordingly, for CHPs in between 7.39 bn €/yr and 9.34 bn €/yr, and the heat pumps in between 16.28 bn €/yr and 19.44 bn €/yr. Therefore, the total cost for the heat generators in Germany varies between 23.67 bn €/yr and 28.78 bn €/yr, which sums up to 316–385 bn € over the lifetime.

For storages, while partial investment is suggested for ACAES varying from 0.15 GW$_{el}$ to 0.51 GW$_{el}$, batteries are preferred over Hydrogen storages for all three scenarios. Existing PHS and ACAES capacities are used throughout the year, and the TES are utilized to their maximum exogenous capacities. The total cost of power and storage shows that maximum investment is required for Li-ion batteries (1.57 bn €/yr) and TES storage capacities (1.1–2.21 bn €/yr). With minimum storage provision in the conservative scenario and maximum storage provision in the progressive scenario, the total cost for electrical and heat storages in Germany varies between 2.67 bn €/yr (conservative) and 3.86 bn €/yr (progressive), which sums up to around 33–48 bn € over the full lifetime.

The energy mix comparison with Fraunhofer ISE's studies suggests that a transformation towards 100%-system according to the OSeEM-DE model results is feasible by 2050, where the energy mix consists of onshore wind, solar PV, offshore wind, CHP, and hydro ROR plants. The biomass-CHP is a promising option to replace the 15% fossil-fuel-based generation of the 85% system of Fraunhofer ISE's study, and the optimized results from the OSeEM-DE model shows that the German potential is sufficient to meet this requirement. The LCOE values have been compared with another Fraunhofer ISE study, which cross-validates the model results.

The flexibility of the system was examined via sensitivity analysis of the grid and storage capacities. It is observed that with grid expansion, the cost of offshore wind, solar PV, and heat pump decreases, but the CHP investment increases slightly. On the other hand, with the gradual expansion of storage, offshore wind, CHP, heat pump investment, and energy transfer from NDE to SDE decreases. Therefore, maximum utilization of the storage usage and optimum grid expansion can provide additional flexibility to the system and decrease the overall investment cost. The cost of grid expansion vs. investment in generation and storage facilities must be investigated to reach the optimum solution. The limited capacity of battery materials should also be taken into consideration. A possible alternative storage solution is Norwegian hydro storages with the interconnection between Norway and Germany, subject to further investigation.

*7.2. Limitations of the study*

This study's modeling approach has five main aspects that could alter the results:

1. Detailed transmission modeling and consideration of transmission costs;
2. Inclusion of industrial heating demand;
3. Inclusion of other renewable and heating technologies;
4. Inclusion of other storage options; and
5. Interconnection with neighboring countries.

Regarding the transmission modeling, the transmission lines between Northern and Southern Germany were modeled as transshipment capacities between two nodes. The internal transmission constraints are not considered, and the grid expansion's investment costs are not calculated. Therefore, the benefits of grid expansion in a sector coupled network in terms of net system cost, including the cost of transmission, could not be determined. Schlachtberger et al. showed that there could be a '*compromise grid*' in-between '*today's grid*' and the '*optimal grid*', in a highly renewable European electricity network [108]. While *optimal grid* expansion can be infeasible due to social acceptance issues, the *compromise grid* offer the maximum benefit at an adequate amount of transmission. A similar scenario is expected in the German electricity system, which should be investigated with detailed transmission grid modeling and consider facts such as meshed networks and power handling capabilities [117].

Second, the heating sector does not include industrial heating demand (i.e., process heat). There are two main reasons: (1) model complexity of high-temperature heating applications, and (2) the availability of hourly industrial heating time series data. The industrial heating demand can be satisfied using several renewable-based approaches, including P2H technologies such as high-temperature heat pumps and electrode boilers, green Hydrogen for cement, iron, and steel production [118], and biogas-based heating. An alternative solution should be figured out in the next phase of the research to translate aggregated industrial heating demand data to hourly time series data. With the inclusion of industrial process heating demand, the feasibility of a 100% power-heat coupled system for Germany needs to be re-investigated.

Third, the current model only considers solar PV, wind, hydro, and biomass and does not include other options such as concentrated solar power (CSP) and geothermal plants. The heating is therefore highly dependent upon the use of biomass, which is limited in potential. A model run without biomass resources as heating, using only electricity-based heat pumps, resulted in an infeasible optimization solution. The inclusion of CSP and geothermal plants and other heating technologies such as electric and electrode boilers and green Hydrogen-based heating may reduce the need for biomass-based CHPs, which should be investigated in future research works.

Fourth, only hot water-based TES was considered as a heat storage option in the model. Other storage options, such as long-term (seasonal) underground TES in boreholes or water pits, were not considered in the model. The main reason for excluding these options is the complex modeling characteristics of these types of storage. The model also simplifies the hot water storage based on simplified assumptions, such as the water inside the tank is thoroughly mixed and has a uniform temperature. Such a simplified model can offer simple and computationally feasible solutions for aggregated power systems. Nevertheless, in the case of a comprehensive analysis of energy systems, it is essential to include all the possible heat storage options with detailed modeling.

Finally, the import and export of electricity between countries, an essential part of the combined energy transition towards the EU's climate-neutrality, were not considered in the model. The inclusion of electricity trade with neighboring countries will affect the calculated investment capacities and costs. The gradual inclusion of all the EU countries is currently considered a future development step of the model.

The model results will also alter if it considers different factors such as higher work/energy ratio of electricity over combustion, elimination of upstream emissions, and policy-driven increases in end-use energy efficiency, as described by Jacobson et al. [64]. Besides, this study considered low PHS inflow values (dry season). There will be more PHS provisions for high inflow values (wet season), affecting the storage investment. The least-cost options for a 100% renewable energy system are not limited to the current OSeEM-DE model components. While this study shows one of the possible pathways for achieving decarbonization in power and building heat sectors, the result will differ when other energy sectors are coupled. There can be a different set of solutions with cheaper Hydrogen or a system with vehicle-to-grid (V2G) and demand-side management (DSM). The benefits of highly renewable and sector-coupled systems can only be realized with a combination of different solutions. For example, the concept of 'Smart Energy Systems' includes solid, gas, and liquid fuel storage, heat pumps with TES, battery electric vehicles, smart thermal grids (district heating and cooling), ICT-based smart electricity grids, smart gas grids, and other fuel infrastructures [21–25,31,36,37,71,119,120]. The concept of 'WWS' includes wave, tidal, geothermal, and CSP plants with the other renewable options for providing 100% energy [55,56,59–64,80,81]. Besides, 'WWS' studies also focus on issues such as social costs and job creation. The PyPSA based studies focus on the benefits of sector-coupling and transmission expansion in a highly interconnected and renewable-based European network [95,106–109]. The author refers to all the relevant '100% system' and 'Sector Coupling' studies since there is not a single pathway for the energy transition, but many possible pathways, with different advantages and disadvantages. The OSeEM-DE model should be seen as a tool, which can be adjusted for different scale energy systems for the EU to investigate different energy-related research questions.

The author wants to clarify that the mentioned limitations do not affect the current model's architecture. Only the results will alter with the inclusion of new technologies (e.g., transmission line, CSP, geothermal, electrode boiler, H2-heating), new costs (e.g., transmission expansion cost), new demand data (e.g., industrial heating, transport), and interconnections (e.g., between Germany and Norway). The methodology, including mathematical formulations as described in Section 3, and the development of the architecture as described in Section 4, will not change in further development. Nevertheless, the model upgrade plans, which are discussed in detail in Section 7.3, will allow the user to investigate more scenarios. For example, considering the inclusion of detailed transmission line modeling, the second version of the model (e.g., OSeEM-DE v2.0) will enable a user to investigate the case of grid expansion against sector-coupling, and the results will be different from this study. However, the previous architecture remains the same, and the only change, in this case, will be the addition of transmission model components and respective system costs to the first version of the model.

*7.3. Future steps of the study*

The open modeling tool OSeEM-DE paves the pathway towards modeling and analyzing plausible sector-coupled scenarios for 100% renewable-based national and sub-national energy systems. At the same time, this study shows how different energy mix options and their component-wise investment capacities and costs can be investigated using the model; hence, Germany's case can be followed for other similar regions to conduct the feasibility analysis of 100% renewable-based sector-coupled systems. The study also reveals that sensitivity analyses can help identify the system's flexibility aspects in future energy infrastructure. The following modeling plans are outlined for further development–

1. Detailed transmission grid and high voltage direct current (HVDC) transmission line;
2. High-temperature industrial process heating and district heating network components;
3. Renewable technology components such as CSP, geothermal plants, and solar thermal collectors;
4. Storage components such as latent heat and chemical heat storages;
5. Transport components such as battery and fuel cell electric vehicles with the provision of V2G; and
6. DSM components.

The model results will be compared and cross-validated with similar modeling tools as described in [3]. The model will be regularly updated in GitHub [121], with source code and input data, so that other energy researchers can use the model for investigating different research questions for the renewable and sector-coupled energy systems in the EU context. According to the current study plan, the author plans to investigate the following research questions–

1. Feasibility of a 100% renewable power-heat-transport coupled energy system for Germany;
2. Impact of Nordic hydro expansion on electricity cost and supply mix for the European energy system in 2050;
3. Investigation on Nordic countries' profitability as flexibility providers for the highly interconnected continental Europe using their green batteries.

In this study, the maximum capacity density on the available area is considered to be 4 MW/km$^2$ for onshore wind and 6 MW/km$^2$ for offshore wind plants, according to the *LIMES-EU* project [104]. Similar strategy to consider land limitations were also considered in studies by Brown et al. [106], Schlachtberger et al. [107,108], and Hörsch et al. [109]. A recent study from Enevoldsen and Jacobson estimates that the maximum output power densities are much higher for Europe, 19.8 MW/km$^2$ for onshore wind farms, and 7.2 MW/km$^2$ for offshore wind farms [122]. The updated values in output power densities directly impact the large-scale development of onshore wind power plants since significantly fewer land areas will be needed for new wind project developments. [122] also implies that the electrical infrastructure and the land acquisition costs will be lowered. Therefore, with the updated capacity density values from [122] into consideration, OSeEM-DE will have a higher value of maximum available potential as the optimization model inputs. This is particularly important for future investigations of OSeEM-DE, with industrial heat and transport demands in the model. The increased maximum potential will allow the installation of more wind turbines, with reduced investment costs, to satisfy the industrial heat and transport loads, providing additional flexibility to decarbonize the industrial heat and transport sectors.

*7.4. Novelty and usability of the model*

The OSeEM-DE is a comprehensive energy system *model* constructed using the toolbox from the *framework* Oemof, which includes the *model generator* Oemof Solph. In [123], Hilpert et al. clarified the definitions and relationship between *model*, *model generator*, and *framework*. The *models* represent the real-world energy systems with a specific regional focus and temporal resolution concretely and may consist of linked sub-models to answer straightforward research questions. *Model generators* employ specific analytical and mathematical approaches by using predefined sets of equations. A *framework* is a structured toolbox that includes sub-frameworks, model generators, and specific models. Besides, an *application* can be developed using one or more *framework* libraries depending on scope and purpose. The Oemof libraries can be used to build different energy system models, which can be referred to as *applications*. Therefore, according to the definition from [123], OSeEM-DE is a model that represents the energy system of Germany with an hourly temporal resolution. The model is novel and unique for the following reasons-

1. A unique feature of the OSeEM-DE model is that it allows a user to modify the model at two different levels. At the basic level, the user can use the model with simple tabular data and scripts. Therefore, the basic level user can visualize the scenario beforehand, prepare the scenario by changing input data in *.csv* data files to develop the scenarios, use the given Python scripts and modify them to reflect his developed scenario. On the other hand, the advanced level users can access the underlying structures, including the *model generator* Oemof Solph and the *class* Oemof Tabular *Facades* to enhance the analytical and mathematical formulations. The unique feature allows energy system analysis ranging from aggregated analysis (e. g., basic level analysis with low resolution) to comprehensive analysis (e.g., advanced level analysis of energy systems with high resolution)
2. The model follows a hybrid approach where the users can define current capacities exogenously, and the future investments are determined endogenously, with a limit of the maximum potential of different technologies. The unique feature allows to create different scenarios ranging from brownfield to greenfield approaches and evaluate a more holistic view of the energy system investments.
3. The use of heat pumps with storage as dispatchable loads is another unique feature of the model. OSeEM-DE uses heat pump technologies (ASHP and GSHP) combined with hot water-based heat storage (TES), which allows flexible operation of heat pumps in 100% renewable energy systems and reduces the need for electricity storage units.
4. While a previous model (REMod-D) [29,30] represented 60% of the building heat demand, OSeEM-DE presents 100% of the demand to decarbonize the building heat sector of Germany fully. Besides, OSeEM-DE is an open-source model in contrast to other models doing similar investigations.
5. The model offers extended storage provisions. The model considers Redox, $H_2$, and ACAES as storage options in addition to traditional Li-ion and PHS options. The use of ACAES is unique because it is not very common to include the technology in energy models. The availability of salt caverns in Northern Germany allows a considerable potential for decentralized renewable energy storage, which is considered an essential input of the OSeEM-DE model.
6. The model considers biomass potential from forest residues, livestock effluents, and agricultural residues in Germany to use them as CHP fuels. In contrast to energy models, which rely heavily on P2H for heat decarbonization, the OSeEM-DE model uses biomass-CHPs with realistic biomass availability data from the *Hotmaps* project [111]. Using a combination of P2H and traditional CHPs based on renewable resources is unique and new and paves the pathway for similar future analysis.
7. The model is the first to analyze the interaction between Northern and Southern Germany to reflect on the flexibility aspects of grid expansions. A similar method can be followed for analyzing import/export in other national energy systems.

8. Conclusion

One of the objectives for developing the model is to use simple scripts and tabular data sources to construct complex energy systems. For this purpose, the author used Oemof, one of the 16 tools selected based on modeling 100% renewable and sector-coupled systems in the NS region. The underlying concept is the Oemof Solph package, which is based on a graph structure at its core and provides an optimization model generator to build investment and dispatch models. While the internal architecture of Oemof Solph is comparatively complex and intended for the expert developers, the OSeEM-DE model is intended for the users who can use existing Oemof Tabular *Facades*, tabular *.csv* data files, and simple Python scripts to build the energy system. The current *Facades* include dispatchable generation (to allow modeling of fossil-fuel-based generation), volatile generation, storage, reservoir (for pumped hydro storage), backpressure and extraction turbines (for CHP), commodity (for limiting the amount of available fuel), conversion (for modeling transformers or heat pumps), load, link (for transshipment-based transmission), and excess (for handling the excess generation from the renewable sources in the optimization model). These classes can also be mixed with the Oemof Solph classes, which broadens the model's versatility. The implementation of these components in Oemof Solph is transparent, and the equations and logics behind the component development are also available in the documentation of Oemof Solph.

To use the OSeEM-DE model for different energy system contexts, the user needs to add or delete the relevant components, buses and substitute the input data. The simple scripting method also allows splitting up countries into several regions, which allows for comprehensive subregional analysis. For example, suppose a user wants to investigate the impacts of connecting the Norwegian pumped hydro storage to the German energy system. In that case, he can develop the OSeEM-DENO model adding the Norwegian energy system using similar scripts but different input data. In any generation that differs from the German system, the user can use the relevant *Facade* class to create the component and add it to the model. If there is a nuclear generation in the system, the user can write a simple script for adding the 'dispatchable' *Facade* class with relevant nuclear input data in the tabular file. Similarly, the model can scale up to a highly interconnected European energy system (e.g., OSeEM-EU), where each country is represented by one or several nodes. The availability of the source code and input data of the model increases transparency and reduces the user's effort to build up a different energy system for investigation.

**Data:** Data, source codes, and results of the model is available in the following Github repository [121]: https://github.com/znes/OSeEM-DE.

**Funding**

This research is a part of the ENSYSTRA project which received funding from the European Union's Horizon 2020 research and innovation programme under the Marie Skłodowska-Curie grant agreement No: 765515.

**Declaration of Competing Interest**

The author declares that he has no known competing financial interests or personal relationships that could have appeared to influence the work reported in this paper.

**Acknowledgements**

This work was supported by the ENSYSTRA network and by the Center for Sustainable Energy Systems (ZNES) in Europa-Universität Flensburg. The author is grateful to receive regular supervision and guidance from Prof. Dr. Olav Hohmeyer during the ENSYSTRA project. The author is also indebted to Andre Harewood for reviewing the paper's language. The author is thankful to Simon Hilpert, Christian Fleischer, and Martin Söthe for answering his last-minute questions. The author appreciates the consistent support from his wife, Anika, during the Covid-19 pandemic's difficult days.

**Appendix A**

The input data for cost calculation are obtained from various resources [93,96–113,115,124] as summarized in Table A.1. All data are available in the Github repository [121].

**Table A.1**
Input Cost Data for OSeEM-DE Model.

| Technology | Onshore Wind | Offshore Wind | Solar PV | Hydro (ROR) | Biomass | Li-ion | $H_2$ | Redox | PHS | ASHP | GSHP | ACAES | TES |
|---|---|---|---|---|---|---|---|---|---|---|---|---|---|
| Capex (€/kW) | 1075 | 2093 | 425 | 3000 | 1951 | 35 | 1000 | 600 | 2000 | 1050 | 1400 | 750 | 0 |
| Lifetime (Years) | 25 | 25 | 25 | 50 | 30 | 20 | 22.5 | 25 | 50 | 20 | 20 | 30 | 20 |
| WACC | 0.025 | 0.048 | 0.021 | 0.05 | 0.05 | 0.05 | 0.05 | 0.05 | 0.05 | 0.05 | 0.05 | 0.05 | 0.05 |
| VOM Cost (€/MWh) | 0 | 0 | 0 | 0 | 11.3 | 1 | 1 | 1 | 0 | 0 | 0 | 1 | 0 |
| FOM Cost (€/kWh) | 35 | 80 | 25 | 60 | 100 | 10 | 10 | 10 | 20 | 36.75 | 49 | 10 | 0.38 |
| Storage Capacity Cost (€/kWh) | – | – | – | – | – | 187 | 0.2 | 70 | – | – | – | 40 | 38 |
| Carrier Cost (€/MWh) | – | – | – | – | 34.89 | – | – | – | – | – | – | – | – |

**9. Broader Context**

The study contributes to open energy modeling challenges to analyze 100% renewable-based energy systems for integrated energy sectors. Additionally, it contributes to renewable energy utilization studies and the understanding of energy transition in the German context. The research illustrates an in-depth methodology for developing renewable and cross-sectoral energy system models and delineates a scenario-based energy system analysis technique. It advances the understanding of the integrated system's feasibility and shows the required capacities and investment for its components to stay within renewable resources' available potentials. The study reveals the possibility of using power-to-heat devices such as electric heat pumps and using biomass-based combined heat and power plants to eliminate the dependency on fossil fuels for heat generation. The study also shows that grid and storage expansion impact the energy system's flexibility. To summarize, the model and the analysis technique will be useful for investigating similar energy systems, especially in the EU context.

**Appendix B. Supplementary data**

Supplementary data to this article can be found online at https://doi.org/10.1016/j.apenergy.2021.116618.